\documentclass[aps,prd,twocolumn,twoside,superscriptaddress,nofootinbib,longbibliography,preprintnumbers,bibnotes]{revtex4-1}
\PassOptionsToPackage{table,xcdraw}{xcolor}

\usepackage{hyperref}
\usepackage{graphicx}
\usepackage{amssymb}
\usepackage{amsmath}
\usepackage{mathrsfs}
\usepackage{xspace}
\usepackage{xcolor}
\usepackage{tikz}

\usepackage[caption=false, labelformat=simple, font={bf}]{subfig}

\usepackage{booktabs}
\usepackage{array} 

\usepackage{algorithm}
\usepackage{algpseudocode}
\definecolor{codegreen}{rgb}{0,0.6,0}

\usepackage{marginnote}
\usepackage[normalem]{ulem}

\definecolor{ao(english)}{rgb}{0.0, 0.5, 0.0}

\newcommand{\eq}[1]{Eq.~\eqref{eq:#1}}
\newcommand{\eqs}[2]{Eqs.~\eqref{eq:#1} and \eqref{eq:#2}}

\newcommand{\fig}[1]{Fig.~\ref{fig:#1}}
\newcommand{\figs}[2]{Figs.~\ref{fig:#1} and \ref{fig:#2}}
\newcommand{\figss}[3]{Figs.~\ref{fig:#1},~\ref{fig:#2}, and~\ref{fig:#3}}
\newcommand{\app}[1]{App.~\ref{app:#1}}

\newcommand{\ord}[1]{\mathcal{O}(#1)}

\newcommand{\df}{\mathrm{d}}

\newcommand{\al}{\alpha}

\newcommand{\de}{\delta}

\newcommand{\si}{\sigma}

\newcommand{\nn}{\nonumber}

\newif\ifrecursive

\newcommand{\izero}{\ensuremath{s}}

\allowdisplaybreaks[2]

\begin{document}


\title{New Angles on Energy Correlators}

\author{Samuel Alipour-fard}
\email{samuelaf@mit.edu}
\affiliation{Center for Theoretical Physics, Massachusetts Institute of Technology, Cambridge, MA 02139, USA}

\author{Ankita Budhraja}
\email{abudhraj@nikhef.nl}
\affiliation{Nikhef, Theory Group,
	Science Park 105, 1098 XG, Amsterdam, The Netherlands}

\author{Jesse Thaler}
\email{jthaler@mit.edu}
\affiliation{Center for Theoretical Physics, Massachusetts Institute of Technology, Cambridge, MA 02139, USA}

\author{Wouter J.~Waalewijn}
\email{w.j.waalewijn@uva.nl}
\affiliation{Nikhef, Theory Group,
	Science Park 105, 1098 XG, Amsterdam, The Netherlands}
\affiliation{Institute for Theoretical Physics Amsterdam and Delta Institute for Theoretical Physics, University of Amsterdam, Science Park 904, 1098 XH Amsterdam, The Netherlands}

\preprint{MIT-CTP/5794}

\begin{abstract}
Energy correlators have recently come to the forefront of jet substructure studies at colliders due to their remarkable properties:
they naturally separate physics at different scales, are robust to contamination from soft radiation, and offer a direct connection with quantum field theory.
The current parametrization used for energy correlators, however, is based on redundant pairwise angles with complex phase space restrictions.
In this Letter, we introduce a new parametrization of energy correlators that features a simpler phase space structure and preserves information about the orientation of jet constituents.
Further, our parametrization drastically reduces the computational cost to compute energy correlators on experimental data; whereas the time to compute a traditional projected $N$-point energy correlator scales as $M^N/N!$ on a jet with $M$ particles, our new parametrization achieves a scaling of $M^2 \ln M$, remarkably independently of $N$.
Even for \(N=3\), this improved scaling is particularly important for studies of heavy ion collisions, and higher values of \(N\) will enable new qualitative understanding of gauge theories.
Theoretical calculations for our new energy correlators differ from those of traditional parametrizations only at next-to-next-to-leading logarithmic accuracy and beyond, and we expect that our simpler phase space structure will simplify those calculations.
We also discuss how to extend our parametrization to resolved $N$-point energy correlators that encode angular distances between greater numbers of particles, yielding intuitive visualizations of jet substructure that are qualitatively different for different jet samples.
We propose two possible generalizations for probing multi-prong jets and testing jet scaling behavior.
\end{abstract}


\maketitle

\emph{Introduction}
--- 
%
The flow of energy within hadronic jets is an indispensable probe of Quantum Chromodynamics (QCD)~\cite{PhysRevD.1.1416, PhysRevLett.39.1237, PhysRevLett.39.1587, PARISI197865, PhysRevD.20.2759, RAKOW198163}.
Energy correlator observables~\cite{Sterman:1975xv, Basham:1977iq, Basham:1978bw, Basham:1978zq, Basham:1979gh} are particularly powerful tools for understanding energy flow both theoretically and experimentally~\cite{Mazzilli:2024ots,CMS:2024mlf,Tamis:2023guc}.
Since energy correlators can be described directly in terms of field-theoretic energy flow operators~\cite{Sveshnikov:1995vi, Tkachov:1995kk, Korchemsky:1999kt, Lee:2006nr, Hofman:2008ar, Belitsky:2013xxa, Belitsky:2013bja, Kravchuk:2018htv}, one can use sophisticated theoretical techniques, including the powerful technology of conformal field theories~\cite{Hofman:2008ar, Kologlu:2019mfz}, to extract rich information about jet substructure, especially in the collinear limit~\cite{Dixon:2019uzg, Chen:2020vvp, Chen:2019bpb, Chen:2020adz, Chen:2021gdk, Schindler:2023cww, Gao:2023ivm, Chen:2023zlx, Chicherin:2024ifn, Chen:2024nyc, Chang:2022ryc,Budhraja:2024xiq,Budhraja:2024tev,Lee:2024jnt}.

Recent work has highlighted the role of $N$-point energy correlators (ENCs) in precisely understanding the fundamental structure of particle interactions.
ENCs probe angular correlations between $N$ final-state particles, which offers a simple and intuitive way to separate physics at different scales and mitigate contamination from soft radiation.
Applications focused on the Large Hadron Collider (LHC) include the top quark mass \cite{Holguin:2022epo, Holguin:2023bjf, Holguin:2024tkz}, hadronization transition~\cite{Komiske:2022enw, Lee:2024esz}, dead-cone effect~\cite{Craft:2022kdo}, medium modifications in heavy-ion collisions~\cite{Andres:2022ovj, Andres:2023xwr, Barata:2023zqg, Barata:2023bhh, Andres:2023ymw, Singh:2024vwb, CMS:2024ovv, Bossi:2024qho}, and predictions for the energy flow of charged particles~\cite{Li:2021zcf, Chen:2022muj, Chen:2022pdu, Jaarsma:2023ell}.
Energy correlators have also been used in studies of gluon saturation and nuclear tomography  \cite{Karapetyan:2019fst,Liu:2022wop,Liu:2023aqb,Kang:2023gvg,Cao:2023oef}.
Further, energy correlators have produced the most precise jet substructure measurement of the strong coupling constant to date~\cite{CMS:2024mlf}. 

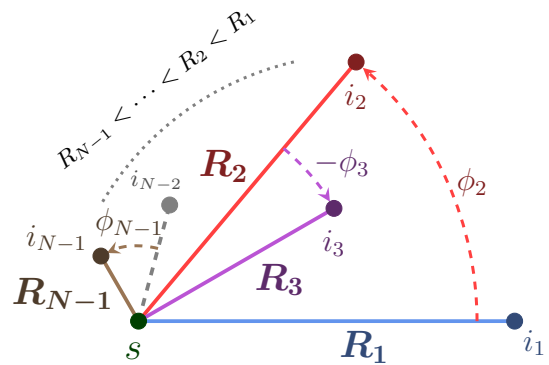
\begin{figure}
    \recursivetrue
	\begin{tikzpicture}
\definecolor{cornflowerblue}{rgb}{0.39, 0.58, 0.93}
\definecolor{azure(colorwheel)}{rgb}{0.0, 0.5, 1.0}

\definecolor{coralred}{rgb}{1.0, 0.25, 0.25}
\definecolor{cadmiumorange}{rgb}{0.93, 0.53, 0.18}
\definecolor{darkgoldenrod}{rgb}{0.72, 0.53, 0.04}

\definecolor{rosevale}{rgb}{0.67, 0.31, 0.32}
\definecolor{palebrown}{rgb}{0.6, 0.46, 0.33}

\definecolor{mediumorchid}{rgb}{0.73, 0.33, 0.83}

\definecolor{ao}{rgb}{0.0, 0.5, 0.0}
\definecolor{lightseagreen}{rgb}{0.13, 0.7, 0.67}

\colorlet{colsp}{ao}
\colorlet{coli1}{cornflowerblue}
\colorlet{coli2}{coralred}
\colorlet{coli3}{mediumorchid}
\colorlet{coliN}{palebrown}


\draw[color=coliN, line width=0.5mm]
    (0, 0) -- (120:1.0) coordinate (iN)
    node[pos=0.4, left, text=black, font=\Large]
    {\textcolor{coliN!50!black}{$\boldsymbol{R_{N-1}}$}};

\ifrecursive
\draw[color=gray, dashed, line width=0.5mm]
    (0, 0) -- (75:1.6) coordinate (iNm)
    node[pos=0.4, left, text=black, font=\Large] {};
\filldraw[color=gray] 
    (iNm) circle (3pt)
    node[above, xshift=-5pt, yshift=2pt]
    {\textcolor{gray!60!black}{$i_{N-2}$}};
\draw[-stealth, color=coliN, dashed, line width=0.4mm]
    (75:1.0) 
    arc[start angle=75, end angle=115, radius=1.0]
    node[pos=0.55, above, text=black, font=\large]
    {\textcolor{coliN!50!black}{$\phi_{N-1}$}};
\else
\draw[-stealth, color=coliN, dashed, line width=0.4mm]
    (0:1.0) 
    arc[start angle=0, end angle=115, radius=1.0]
    node[pos=0.55, above, text=black, font=\large]
    {\textcolor{coliN!50!black}{$\phi_{N-1}$}};
\fi

\filldraw[color=coliN!50!black] 
    (iN) circle (3pt)
    node[above left, xshift=-2pt, yshift=-2pt, font=\large]
    {\textcolor{coliN!50!black}{$i_{N-1}$}};

\draw[color=coli1, line width=0.5mm] 
    (0, 0) -- (0:5) coordinate (i1)
    node[pos=0.6, below, sloped, text=black,
         font=\Large]
        {\textcolor{coli1!50!black}{$\boldsymbol{R_1}$}};
    
\filldraw[color=coli1!50!black] 
    (i1) circle (3pt) 
    node[below right, font=\large]
    {\textcolor{coli1!50!black}{$i_1$}};

\draw[color=coli2, line width=0.5mm]
    (0, 0) -- (50:4.5) coordinate (i2)
    node[pos=0.6, left, text=black, xshift=-4pt, font=\Large]
    {\textcolor{coli2!50!black}{$\boldsymbol{R_2}$}};

\draw[-stealth, color=coli2, dashed, line width=0.4mm] 
    (0:4.5) 
    arc [start angle=0, end angle=48.5, radius=4.5]
    node[pos=0.5, right, text=black, font=\large]
    {\textcolor{coli2!50!black}{$\phi_2$}};

\filldraw[coli2!50!black] 
    (i2) circle (3pt)
    node[below, yshift=-5pt, font=\large]
    {\textcolor{coli2!50!black}{$i_2$}};

\draw[color=coli3, line width=0.5mm]
    (0, 0) -- (30:3) coordinate (i3)
    node[pos=0.6, below, text=black, xshift=9pt, font=\Large]
    {\textcolor{coli3!50!black}{$\boldsymbol{R_3}$}};

\ifrecursive
\draw[-stealth, color=coli3, dashed, line width=0.4mm] 
    (50:3.0) 
    arc [start angle=50, end angle=32, radius=3.0]
    node[pos=0.3, right, xshift=2pt, text=black, font=\large]
    {\textcolor{coli3!50!black}{$-\phi_3$}};
\else
\draw[-stealth, color=coli3, dashed, line width=0.4mm] 
    (0:3.0) 
    arc [start angle=0, end angle=28, radius=3.0]
    node[pos=0.5, right, text=black, font=\large]
    {\textcolor{coli3!50!black}{$-\phi_3$}};
\fi

\filldraw[color=coli3!50!black] 
    (i3) circle (3pt)
    node[below, yshift=-5pt, font=\large]
    {\textcolor{coli3!50!black}{$i_3$}};

\filldraw[color=colsp!50!black]
    (0, 0) circle (3pt) 
    node[below, yshift=-5pt, xshift=-2pt, font=\Large]
    {\textcolor{colsp!50!black}{$s$}};

\draw[gray, dotted, line width=0.4mm,
      xshift=-0.00cm, yshift=0.50cm]
      (55:3.6) 
    arc[start angle=103, end angle=150, radius=4.]
    node[pos=0.55, sloped, above,
         xshift=0.1cm, yshift=0.2cm, text=black]
        {$R_{N-1} < \cdots < R_2 < R_1$};

\end{tikzpicture}
    \caption{
        An illustrative diagram of the new parametrization of ENCs we introduce in \eqs{new_def}{renc}.
        Instead of computing the ENC using all \(\binom{N}{2}\) pairwise distances, we parametrize the ENC with \(2N - 3\) oriented polar coordinates centered on a special particle \izero{}, and then perform a momentum-weighted sum over all choices for \izero{}.
    }
	\label{fig:cartoon}
\end{figure}

In this Letter, we introduce a new parametrization for energy correlators with a number of improved properties.
First, our parametrization of the projected $N$-point energy correlator (PENC) depends on the largest distance \(R_1\) to a ``special'' particle $s$ in a set of $N$ particles, suitably averaged over all choices for $s$;
this yields simpler phase space restrictions than the traditional parametrization for the PENC in terms of the largest pairwise angle~\cite{Chen:2020vvp}.
Second, when considering more differential information, our parametrization of resolved ENCs (RENC) employs non-redundant polar coordinates centered around the special particle, as in \fig{cartoon}, yielding intuitive visualizations of jet substructure.
This differs from the traditional approach, which uses over-complete information from the set of all pairwise distances and neglects information about the relative orientation of particles.
Third, our parametrization offers dramatic improvements in computational performance, which is essential for their use in experimental analyses.
Finally, we anticipate that these conceptual and computational improvements will yield simpler theoretical calculations.
The implementation of the PENCs we introduce in this work can be found on GitHub as an update to \textsc{FastEEC}~\cite{FASTEEC}, and of our PENCs and RENCs at \texttt{ResolvedEnergyCorrelators}~\cite{github:RENC}.

\emph{Review of Energy Correlators}
---
In proton-proton collisions -- the focus of this Letter -- PENCs are usually defined via~\cite{Chen:2020vvp}: 
\begin{align}
    \label{eq:old_def}
    \frac{1}{\sigma}\frac{\df \sigma_N}{\df R_L}
   \! =\!
    \biggl\langle 
        \sum_{i_1 \dots i_N}
        z_{i_1} z_{i_2} \dots z_{i_N}
        \delta\big(
            R_L
            \!-\!
            \max_{k,\ell}
            \{R_{i_k,i_\ell}\}
        \big)
   \!\! \biggr\rangle.
\end{align}
Here, $z_{i}= p_{T,i}/\sum_j p_{T,j}$ is the transverse momentum fraction carried by particle $i$ in the jet, $R_{i_k,i_\ell} = \sqrt{(y_{i_k,i_\ell})^2 + (\phi_{i_k,i_\ell})^2}$ 
is the angular separation between particles $i_k$ and $i_\ell$ in the rapidity-azimuth plane, and the angular brackets indicate an expectation value over a sample of many hadronic jets. 
The sum over $\{i_k\}_{k=1}^N$ indicates the sum over all sets of $N$ particles of a jet, and the variable \(R_L\) characterizes the maximum pairwise angular separation between the \(i_k\).

However, the expression of \eq{old_def} is computationally expensive due to the intricate phase space constraint \(R_L = \max_{k,\ell}\{R_{i_k,i_\ell}\}\).
For example, the time required to compute the (integer) PENC scales as $M^N/N!$ for a jet with $M$ particles; when analytically continued to non-integer values of $N$, the PENC suffers  a computational scaling of $2^{2M}$.%
\footnote{
    \textsc{FastEEC}~\cite{Budhraja:2024xiq} achieves a substantial speed up by replacing particles with subjets whose radius is chosen dynamically, depending on the desired level of angular resolution.
    For non-integer $N$, \textsc{FastEEC} furthermore uses a recursive algorithm that reduces the standard $2^{2M}$ scaling down to $M\,2^M$~\cite{Budhraja:2024tev}.
} 
This computational cost impedes several exciting PENC applications.
For example, PENCs have an approximate scaling behavior related to the $N$-th Mellin moment of the DGLAP splitting functions~\cite{Chen:2020vvp};
the analytic continuation of the PENC to non-integer $N$ therefore provides access both to the full splitting functions of QCD and, in the limit $N \to 0$, a unique opportunity to study small-$x$ physics and BFKL dynamics with jets~\cite{Chen:2020vvp, Budhraja:2024tev,Neill:2020bwv}.
At very large values of \(N\), PENCs also encode additional fundamental features of QCD~\cite{Chen:2020vvp,Dai:2024wff}, such as level crossings with twist-4 operators.
Furthermore, in the high-multiplicity environment of heavy ion collisions, even computing PENCs for \(N=3\) has been computationally challenging, limiting their potential in studying medium effects~\cite{Bossi:2024qho}.
Improved computational efficiency is therefore necessary to leverage the full potential of PENCs in the study of realistic data samples.

Higher-point ENCs also yield more detailed shape information about the structure of radiation inside jets.
For example, recent work has leveraged the E3C to propose a new method for extracting the top quark mass from experimental data~\cite{Holguin:2022epo, Holguin:2023bjf, Holguin:2024tkz}.
Notably, for $N>3$ the parametrization of ENCs in terms of the $R_{i_k,i_\ell}$ is over-complete:
the $R_{i_k,i_\ell}$ describe $N \choose 2$ distances, while only $2N-3$ are independent.

\emph{New Angles: the Projected Case}
---
In this Letter, we introduce a new parametrization of PENCs with a simpler phase space structure:
\begin{align}
    \label{eq:new_def}
    \frac{1}{\sigma}\frac{\df \sigma_N}{\df R_1}
    &= 
    \biggl\langle
        \sum_{\izero=1}^M  z_{\izero} \!\!
        \sum_{i_1 \dots i_{N-1}}
       \!\! z_{i_1} \dots z_{i_{N-1}}
        \delta (
            R_1 
            \!-\! 
            \max_j\{R_{\izero,i_j}\}
        ) \!
    \biggr\rangle ,
    \nn \\
    &\equiv \text{PENC}(R_1) \,
    .
\end{align}
%
The sums on \(\izero\) and \(\{i_j\}_{j=1}^{N-1}\) again run over all $M$ particles within a jet.
The crucial simplification is that our PENC is based on a new variable, \(R_1\), that indicates the maximum distance \(\max\{R_{\izero, i_j}\}\) between the \textit{single} particle \(\izero\) and any of the remaining $N-1$ particles \(\{i_j\}_{j=1}^{N-1}\).

\begin{figure}
    \includegraphics[width=0.47\textwidth]{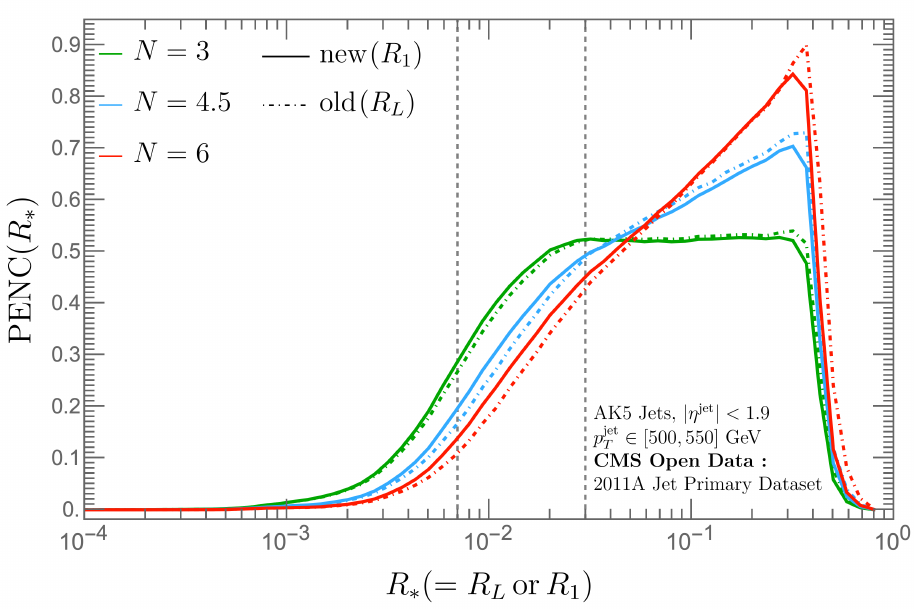} 
    \caption{
        PENC distributions for $N = \{3, 4.5, 6\}$, calculated using our new parametrization (solid) and the traditional parametrization (dashed).
        $R_{*}$ denotes the largest distance to the special particle for our new parametrization ($R_1$), and the largest separation between the $N$ particles for the traditional one ($R_L$).
        The differences are small in the perturbative region $R_{*} \!\! \gg  \! \Lambda_{\rm QCD} /p_T$, but become noticeable in the transition between perturbative and non-perturbative regimes (indicated by the vertical dashed lines).
    }
	\label{fig:ENC}%
\end{figure}

Like the old variable \(R_L\) of \eq{old_def}, \(R_1\) still roughly characterizes the maximum angular scale between a set of \(N\) particles, since \(R_L/2 \leq R_1 \leq R_L\) by the triangle inequality.
Indeed, the PENCs displayed in \fig{ENC} show that the difference between our parametrization and that of \eq{old_def} is small, and that both parametrizations have similar scaling behavior in the perturbative region (though there are some differences in the perturbative to non-perturbative transition).
\figss{ENC}{bullseye}{density} all feature energy correlators evaluated on the CMS 2011A Jet Primary Dataset~\cite{CERNOpenDataPortal, CMS:JetPrimary2011A}, also available in MIT Open Data format~\cite{Komiske:2019jim, komiske_patrick_2019_3340205}, on anti-k$_{\rm t}$ jets with transverse momenta $p_{T}^{\text{jet}} \in [500, 550]$ GeV and pseudorapidity $\vert \eta^{\rm jet} \vert < 1.9$~\cite{Cacciari:2008gp}.

\begin{figure*}
    \centering 
    \subfloat[]{
    	\includegraphics[width=0.33\textwidth]{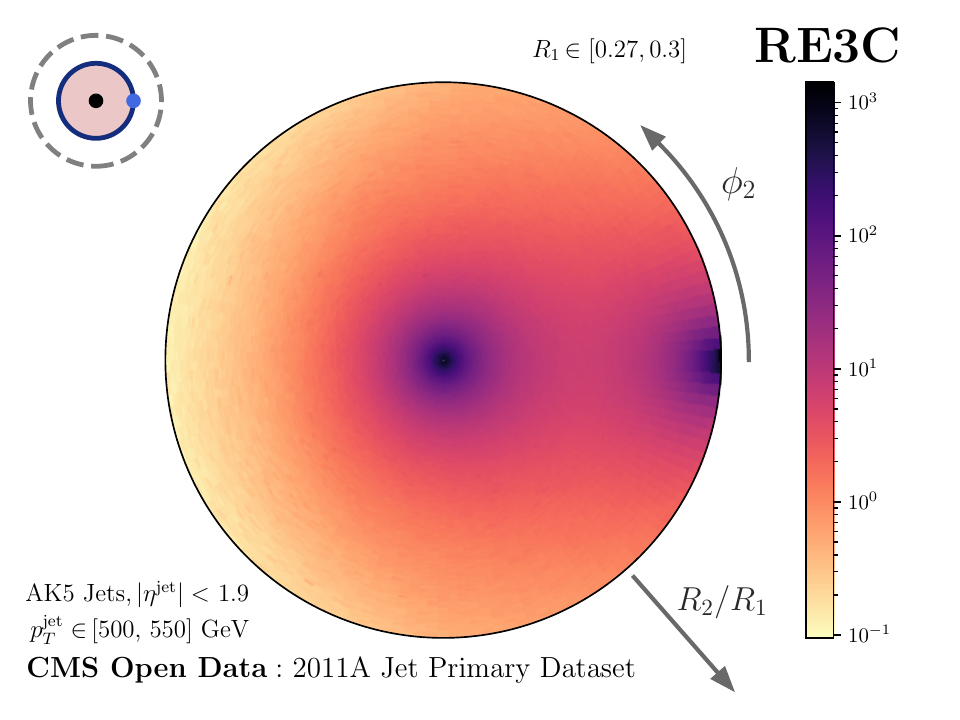} 
        \label{fig:bullseye_new}         
     }
     \subfloat[]{
    	\includegraphics[width=0.33\textwidth]{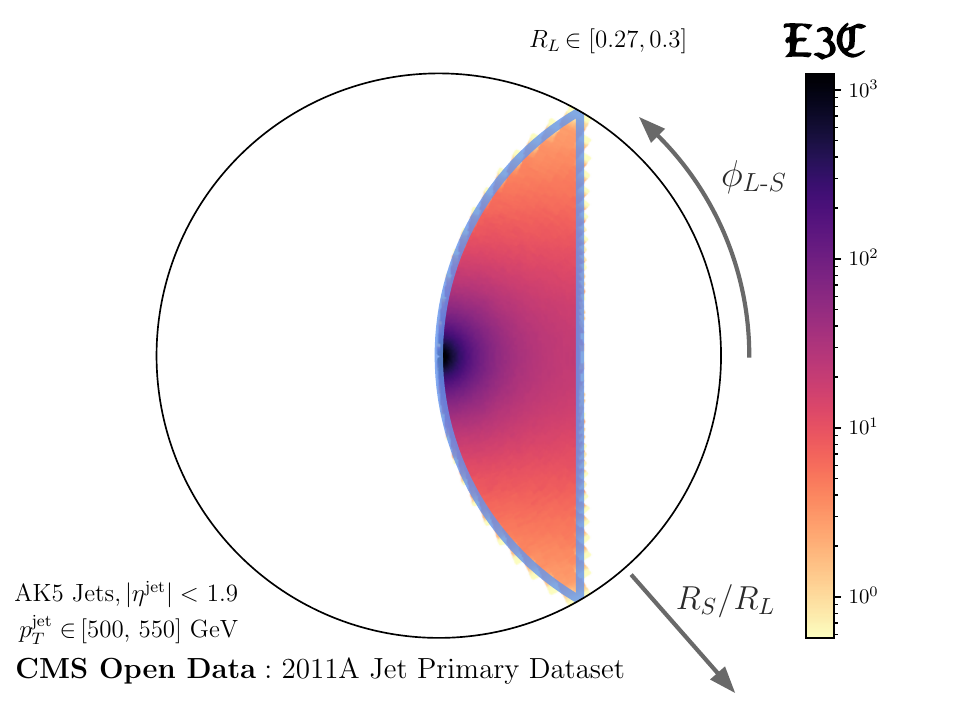}	
        \label{fig:bullseye_old}         
     }
    \subfloat[]{
    	\includegraphics
        [width=0.33\textwidth]
        {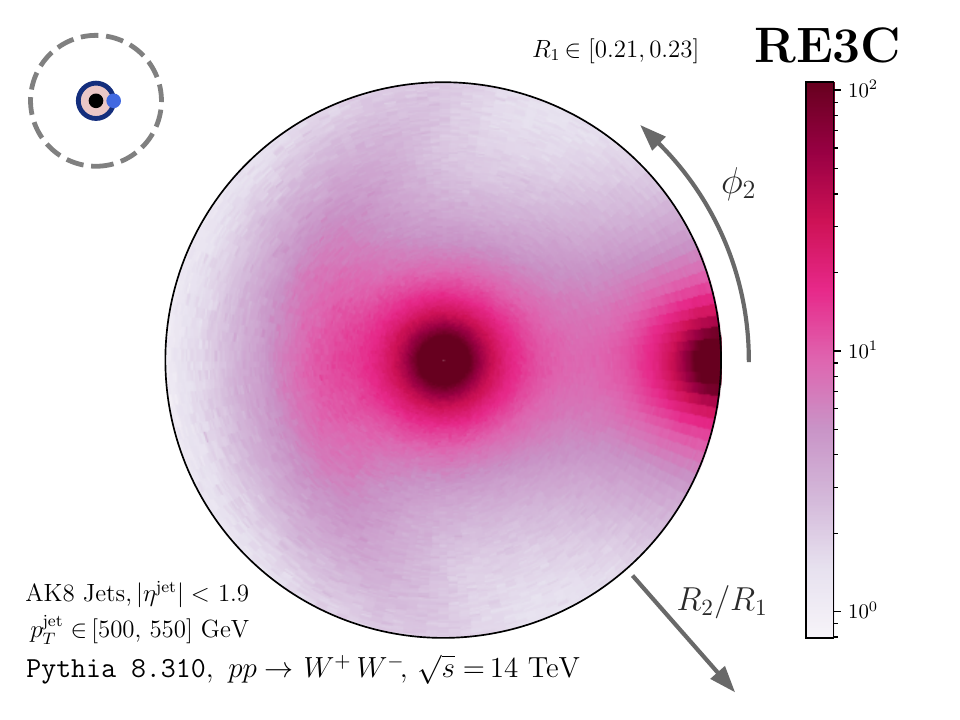}	
        \label{fig:bullseye_new_w}        
    } 
    \caption{
        Polar heat maps visualizing \textbf{(a)} our new RE3C applied to CMS Open Data, \textbf{(b)} the traditional E3C applied to CMS Open Data, and \textbf{(c)} our new parametrization applied to $W$-boson-initiated jets from \texttt{Pythia 8.310}.
        In \textbf{(a)} and \textbf{(c)}, the radial variables correspond to \(R_2/R_1\), the polar angle corresponds to \(\phi_2\), and we show the RE3C in the bin $R_1 \in [0.27,0.3]$.
        In \textbf{(b)} the radius corresponds to \(R_S/R_L\), the polar angle corresponds to the angle between associated lines of length \(R_S\) and \(R_L\), and we show the E3C in the bin $R_L \in [0.27,0.3]$.
        In all three plots, we see collinear enhancements for the RE3C near the origin, when two particles become very close in angle. 
        In \textbf{(a)} and \textbf{(c)}, we also see collinear enhancements as \(R_2/R_1 \to 1\) and \(\phi_2 \to 0\), and \textbf{(c)} exhibits additional non-collinear enhancements correlated with the \(W\)-boson mass.
    }
	\label{fig:bullseye}%
\end{figure*}

\emph{Improved Computation Time}
---
%
The computational efficiency of our parametrization is more evident in the (normalized) cumulative distribution:
\begin{align} \label{eq:new_cml}
  \Sigma_N(R_1)
  \!
  &=
  \!
  \frac{1}{\sigma} 
  \int_0^{R_1}
  \!\! 
  \df R_1'
  \,
  \frac{\df \si_N}{\df R_1'} 
  =
  \biggl\langle \sum_{\izero} z_{\izero} [z_{\rm disk}(\izero,R_1)]^{N-1}
  \biggr\rangle
  ,
\end{align}
where $z_{\rm disk}(\izero,R_1)$ denotes the total transverse momentum fraction of all particles within a radius $R_1$ of the special particle \izero{}.
Notably, the simple form of \eq{new_cml} holds even for non-integer $N$.
A practical way to evaluate \eq{new_cml} (and \eq{new_def} after differentiation) is, for each $\izero$, to first sort all particles by their distance with respect to $\izero$, and then to compute $\Sigma_N(R_1)$ by beginning with $R_1=0$ and then increasing it.
Sorting particles by their distance to \izero{} takes $M\ln M$ time, and the remaining sum over \izero{} scales with $M$, resulting in an overall computation time scaling as $M^2 \ln M$.
This computational speed up is especially interesting for heavy-ion collisions where $M$ is typically very large.
We illustrate the dramatic improvement in the computation time for higher-point energy correlators in the Supplemental Material.

\emph{Theoretical Perspectives}
---
The factorization formula for the traditional PENCs~\cite{Chen:2020vvp} also applies to our new PENCs, with one small difference:
because the variables $R_1$ and $R_L$ differ for three or more emissions, the jet function for our PENCs differs from the old one at $\ord{\alpha_s^2}$, or at next-to-next-to-leading logarithmic accuracy (NNLL).
In the supplemental material, we discuss how the NLL equivalence of jet functions implies that $
    \Sigma_N
    (R_L)
    =
    \Sigma_N(
    R_1 = R_L[
        1 
        +
        \mathcal{O}
        (
        \alpha_s
        )
    ]
    ) 
$%
~\cite{PRLsupplement1}
.
Furthermore, at NNLL and beyond, we expect that the simple dependence of \eq{new_cml} on $N$ will substantially simplify the calculation of the jet function for $R_1$.
By contrast, the jet function using the old parametrization requires dedicated calculations for each individual value of $N$~\cite{Dixon:2019uzg,Chen:2023zlx}.

\begin{figure*}
\centering
    \subfloat[]{
        \includegraphics
        [width=0.33\textwidth]{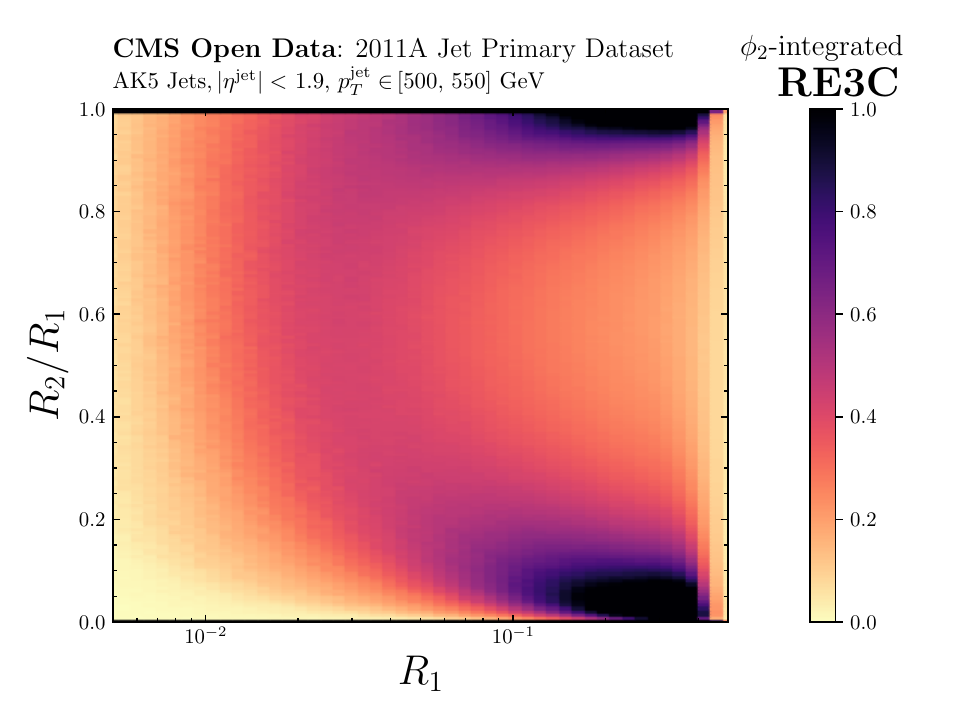}
    	\label{fig:density_new}%
    }
    \subfloat[]{
        \includegraphics
        [width=0.33\textwidth]
        {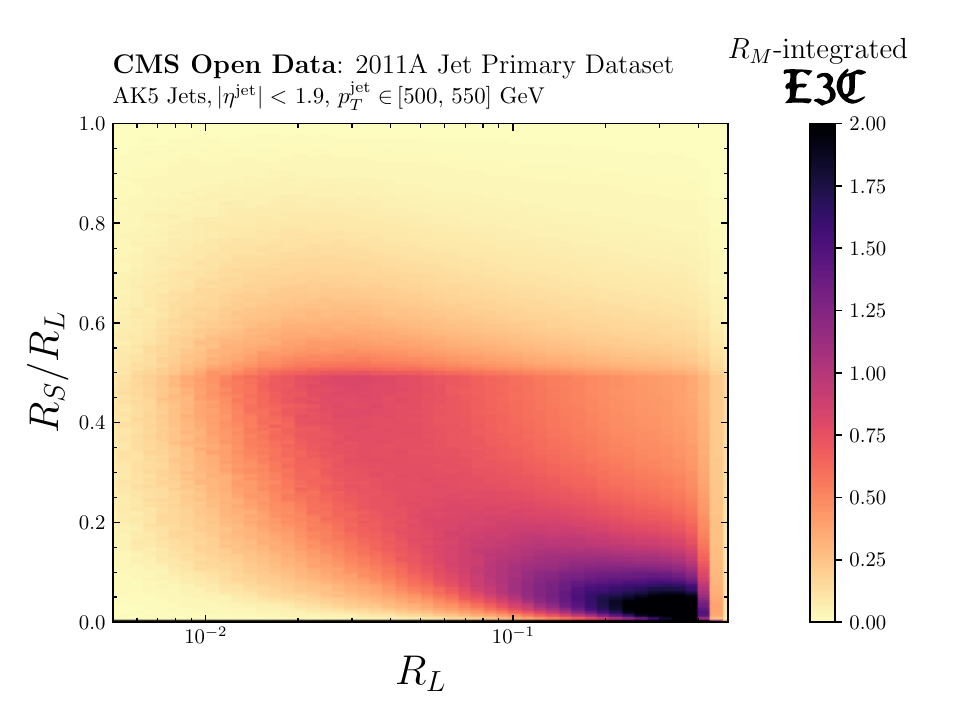}
    	\label{fig:density_old}%
    }    
    \subfloat[]{
        \includegraphics
        [width=0.33\textwidth]
        {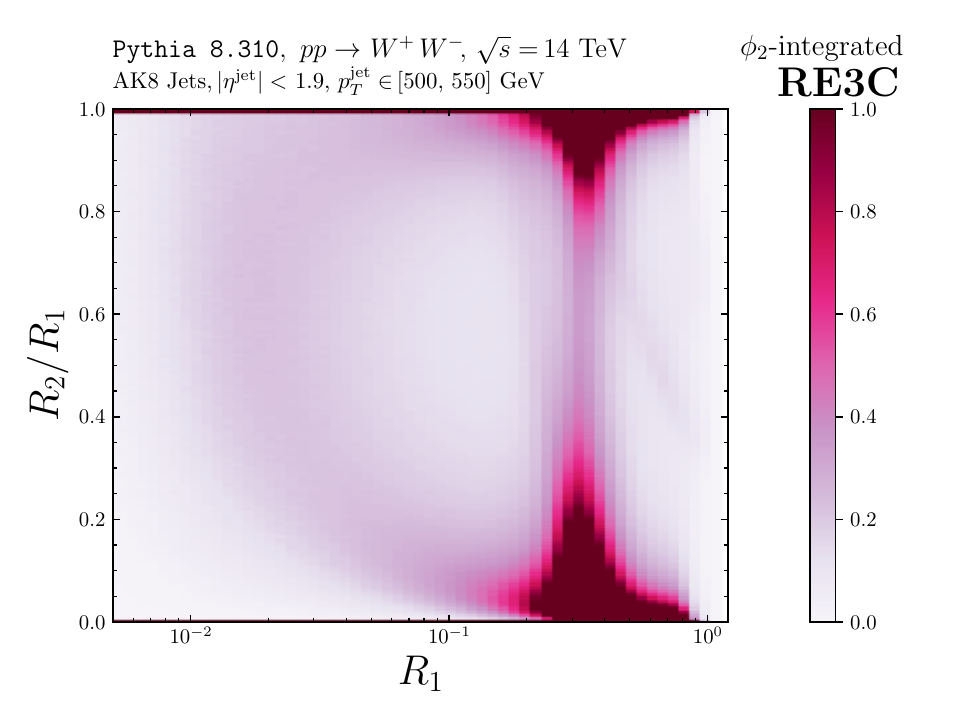}
    	\label{fig:density_w}%
    }
    \caption{
        Normalized distributions \textbf{(a)} our new RE3C applied to CMS Open Data when integrated over the azimuthal angle \(\phi_2\), \textbf{(b)} the traditional E3C applied to CMS Open Data integrated over the intermediate angle $R_M$ (analogous to integrating over \(\phi_s\)), and \textbf{(c)} our new, \(\phi_2\)-integrated RE3C applied to $W$-boson-initiated jets. 
    }
	\label{fig:density}%
\end{figure*}

Finally, we note that \eq{new_cml} is the $N$-th Mellin moment in $z$ of
\begin{align}
    \frac{1}{\sigma}
    \frac{\df \si}{\df z}
    (R_1)
    = 
    \Bigl
    \langle\sum_{\izero} z_{\izero}\, \de
    \bigl[
        z
        - 
        z_{\rm disk}(\izero,R_1)
    \bigr]
    \Bigr\rangle
    \,,
\end{align}
which is also the differential jet rate in a jets-without-jets approach if \(R_1\) is treated as a jet radius~\cite{Bertolini:2013iqa,Bertolini:2015pka}.
A similar moment relation between the jet rate and the original ENC was noted in Ref.~\cite{Lee:2024icn}.

\emph{New Angles: the General Case}
--- 
%
By including (or \ \emph{resolving}) more detailed angular information about the positions and relative orientations of particles within the jet, we may generalize our new parametrization for the PENC to introduce the RENC:
\begin{align}
    \label{eq:renc}
    &\text{RENC}(R_1, R_2, \phi_2, R_3, \phi_3, \dots)
    \equiv \frac{1}{\sigma}
    \frac{\df \sigma_N}{\df R_1 \df R_2 \df \phi_2 \df R_3 \df \phi_3 \dots} \nn
    \\
    & 
    = \biggl\langle
    \sum_{\izero}
    z_{\izero}
    \!\!
    \sum_{i_1 \geq\dots\geq i_{N-1}}
    \!\!\!\!
    z_{i_1} \dots z_{i_{N-1}} 
    \binom{N}{n_1\,n_2\, \dots}\,        
    \de
        (R_1 \!-\! R_{\izero,i_1}
    ) 
    \nn 
    \\ 
    & \qquad\qquad\qquad\qquad 
    \ \times
    \rho_{i_2}(R_2,\phi_2)\, \rho_{i_3}(R_3,\phi_3)\, \dots\biggr\rangle
    \, ,
\end{align}
where for each $s$, the $i_j$ are indexed such that $R_{s,i_1} \ge R_{s,i_2} \ge \ldots$, and the summand involves the per-particle densities: 
\begin{align}
    \rho_{i_j}(R_j, \phi_j) 
    =
    \de(
        R_j 
        - 
        R_{\izero,i_j}
    )\,
    \de(
        \phi_j 
        -
        \phi_{i_{j-1},i_j}
    ) 
\,.\end{align}
The ``$\dots$" correspond to additional  \(R_j\) and \(\phi_j\), and $n_k$ denotes how often particle $k$ of the jet appears among the $i_j$ (such that terms with $n_k>1$ encode self-correlations of particle $k$), with $\sum^M_{k=1} n_k = N$.

Our parametrization is visualized in \fig{cartoon}, and utilizes polar coordinates around $\izero$, ordered in radius $R_1 \! > \! R_2 \! > \! \dots$ and with an \emph{oriented} azimuthal angle $\phi_j$ taken relative to the $(j\!-\!1)$-th resolved emission.
The \(R_j\) and \(\phi_j\) use \(2N-3\) variables to completely characterize the positions of all particles in the jet, relative to the axis defined by particles \(s\) and \(i_1\).
The multinomial coefficient in \eq{renc} arises due to the ordering of the \(R_j\), and accounts for the possibility that two or more of the $i_j$ may be equal.
Integrating inclusively over $\{R_j, \phi_j\}_{j=2}^{N}$ (which sets the third line of \eq{renc} to unity) reduces the RENC to the PENC from \eq{new_def}.

The RENCs we introduce also preserve information about relative orientations, whereas even in the simple example of $N=3$, the traditional E3C parametrization in terms of the largest ($R_L$), medium ($R_M$), and shortest ($R_S$) distances does not.
We visualize the additional orientation information preserved by our RE3Cs in \figs{bullseye_new}{bullseye_old}, where we compare polar heat maps of our parametrization of the RE3C to the traditional paramterization of the E3C.
The additional angular information carried by our new RE3C also leads to striking visual differences when comparing jets initiated by different processes.
For example, the unique characteristics of the RE3C evaluated on $W$-boson initiated jets generated with \texttt{Pythia 8.310}~\cite{Bierlich:2022pfr, Skands:2014pea}, shown in \fig{bullseye_new_w}, clearly distinguish it from the RE3C of QCD-initiated jets.
This qualitative difference is not visible in the traditional parametrization, as shown in the supplemental material together with similar visualizations for additional processes.

The squeezed limit of our new RE3C, \(R_2 \ll R_1\), is quite similar to the squeezed limit of the traditional E3C, \(R_S \ll R_L\);
in this limit, \(R_2 \!\! \sim \!\! R_S\) and \(R_1 \!\! \sim \!\! R_L\).
Our new RE3C and the traditional E3C evaluated on CMS Open Data are compared in \figs{density_new}{density_old}, which demonstrate that they indeed take similar forms when \(R_S / R_L,\,R_2/R_1 < \! \frac{1}{2}\).
When $R_S/R_L \!\! > \!\! \frac{1}{2}$, however, the traditional E3C is suppressed due to the hierarchy $R_S < R_M < R_L$.
In \fig{density_w}, we visualize our new RE3C on \texttt{Pythia}-generated \(W\) jets, which exhibits a non-perturbative ridge similar to the one seen in CMS Open Data as well as additional features at \(R_1 = 0.3\) correlated with the \(W\) mass, \(R_1 \sim 2 m_W / p_T\).
In the supplemental material, we explain the ridge-like features in these plots from the perspective of non-perturbative physics.

Finally, we note that each additional resolved particle (i.e.\ each \((R_j, \phi_j)\) pair) introduces a factor of $M$ to the scaling of the RENC computation time.
In contrast, the computation time for e.g.~the traditional 4-point correlator scales as $M^4$ independent of how many angles were resolved.

\emph{Generalizations}
--- 
For completeness, we mention two additional generalizations of our new parametrization for energy correlators -- RENCs with two special particles and RENCs with two projected angles -- for which we defer more detailed explorations to future work.
For the energy correlator with two special particles, 
\begin{align} 
    \label{eq:two_special}
    &\frac{1}{\sigma} \frac{\df \si_{N,N'}}
    {\df R\, \df R_1\, \df R_1'} 
    =
    \biggl\langle
    \sum_{\izero} 
    z_{\izero}  \sum_{\izero'} 
    z_{\izero'}\, 
    \de(R - R_{\izero\izero'})  
    \\ 
    & \quad \times 
    \sum_{i_1 \dots i_{N-1}}
    z_{i_1} \dots z_{i_{N-1}}
    \delta (R_1 - \max\{R_{\izero,i_k}\})
    \nn
    \\
    &
    \quad  \times 
    \sum_{j_1 \dots j_{N'-1}}
    z_{j_1} \dots z_{j_{N'-1}}
    \delta (R_1' - \max\{R_{\izero',j_k}\})
    \biggr
    \rangle
    \nn\,,
 \end{align}
the two special particles $\izero$ and $\izero'$ are separated by a distance $R$, and $R_1$ and $R_1'$ denote the maximum distance of other particles to $\izero$ and $\izero'$, respectively.
This parametrization may capture interesting features of the radiation patterns of intrinsically 2-prong jets, e.g.~from the hadronic decays of $W$ or Higgs bosons, with a straightforward extension to three or more special particles for higher-prong jets.

We also define a \emph{double}-projected energy-correlator,
\begin{align} 
    \label{eq:double_projected}
    \frac{1}{\sigma} 
    \frac{\df \si_{(a, b)}}{\df R_1\, \df R_2} 
    &= 
    \biggl\langle \sum_{\izero} z_{\izero} 
    \sum_{i_1 \dots i_{a}}
    z_{i_1} \dots z_{i_{a}} 
    \delta(R_1 \!-\! \max\{R_{\izero,i_k}\}) 
    \nn 
    \\
    & \times
    \sum_{j_1 \dots j_{b}}
    z_{j_1} \dots z_{j_{b}} 
    \delta (R_2 - \max\{R_{\izero,j_k}\}) \biggr\rangle
    \,,
\end{align}
where one would typically take $R_1>R_2$.
The associated cumulative distribution, analogous to \eq{new_cml}, is 
\begin{align}
    \label{eq:double_projected_cml}
    & 
    \Sigma_{(a,b)}(R_1,R_2) 
    =
    \frac{1}{\sigma}
    \int_0^{R_1}
    \!
    \df R_1'   \int_0^{R_2}
    \!
    \df R_2'
    \,
    \frac{\df \si_{a,b}}{\df R_1'\, \df R_2'} 
    \nn 
    \\
    & \quad
    = 
    \Bigl\langle \sum_{\izero}
    z_{\izero} 
    [z_{\rm disk}(\izero,R_1)]^a [z_{\rm disk}(\izero,R_2)]^b \Bigr\rangle
    \,,
\end{align}
highlighting that there is no additional complication when $a,b$ are non-integer.
Because the effective scaling of this observable is set by $N = 1+a+b$ and because the $N= 1$ moment of the DGLAP splitting functions vanish, we expect very mild dependence on $R_1$ at fixed $R_2/R_1$ when $b = -a$.
This behavior therefore offers an interesting test of parton shower generators.

\emph{Discussion and Outlook}
--- 
%
Motivated by existing results and the tremendous potential of energy correlators for the characterization of jet substructure, this Letter introduces a new parametrization for higher-point correlators with several distinct advantages for both theoretical and data-driven analyses. 
Our new parametrization exhibits dramatically improved computational efficiency when evaluated on experimental data (without approximations), preserves information about the parity and relative orientation between particles within jets, and has intuitive features that benefit data visualization.
We also expect that the simplified phase space will streamline theoretical calculations, the inclusion of hadronization effects, and pile-up subtraction.
Additionally, our parametrization does not involve any redundancy of phase space variables, unlike traditional parametrizations.
We anticipate that the reparametrization of energy correlators we introduce in this work will benefit current studies of energy flow within hadronic jets and open new avenues for the use of energy correlators in the study of particle collisions.

{\it Acknowledgements}
---
We thank Kyle Lee, Ian Moult, and Aditya Pathak for inspiring discussions.
J.T.\ thanks Arjun Kudinoor, Yen-Jie Lee, and Krishna Rajagopal for conversations about computational and visualization challenges with the traditional ENC parametrization.
A.B.~is supported by the project ``Microscopy of the Quark Gluon Plasma using high-energy probes'' (project number VI.C.182.054) which is partly financed by the Dutch Research Council (NWO). 
S.A.F.\ and J.T.\ are supported by the U.S. DOE Office of High Energy
Physics under grant number DE-SC0012567.
J.T. is additionally supported by the Simons Foundation through Investigator grant 929241.
W.W.~would like to thank the MIT CTP for hospitality, where this work was initiated.

\bibliography{EEC_ref}

\clearpage

\onecolumngrid

\section*{Supplemental Material To \\
  New Angles on Energy Correlators}

\appendix

In this \textit{Supplemental Material}, we provide additional details regarding the parametrization we introduce for projected and resolved $N$-point energy correlators (PENCs and RENCs, respectively).
In \app{nll_equiv}, we discuss and visualize the theoretical similarities between our new parametrization and the standard parametrization for PENCs.
In \app{samp_imp_perf}, we discuss the algorithmic implementation and the performance of our new parametrization.
We also show additional results using Open Data, up to RE4Cs, in \app{opendata}, and visualize the process dependence of our PENCs and RENCs on samples for QCD-, $W$-, and top-initiated jets in \app{pythia}.
Finally, in \app{nonpert}, we discuss the qualitative behavior of non-perturbative effects.

\section{NLL Equivalence to Traditional PENCs}
\label{app:nll_equiv}

We first emphasize that the parametrization we propose for PENCs only differs from the traditional PENCs studied in the literature through the jet function.
In particular, the factorization for our new parametrization is identical to that of the traditional PENCs in the collinear limit \cite{Dixon:2019uzg, Chen:2020vvp}:
\begin{align} \label{eq:fact}
    \Sigma_N
    \left(
    R_1, Q
    \right) 
    = 
    \int_0^1\! 
    \df x\, 
    x^N 
    \sum_i H_i
    \Bigl(x,\frac{Q}{\mu}\Bigr) \cdot
    J_i
    \Bigl( \frac{x^2 Q^2R_1^2}{\mu^2} \Bigr)
    \,,
\end{align}
where the flavor index \(i = q,g\)  is summed over.
At a specific perturbative order, the expressions also depend on $\mu$ through $\alpha_s(\mu)$.
%
\eq{fact} states that the resummed cross section for the PENCs 
is governed by: 
\begin{itemize}
    \item 
    A \textit{hard function} \(H_i(x, Q/\mu)\), characterizing the number density of partons of partonic flavor \(i\) (e.g.\ up quark, gluon) emerging from the hard process at a given momentum fraction \(x\).
    The hard function is dependent on a renormalization scale \(\mu\).
    Notably, the hard function is independent of the parametrization used for the PENC.
    
    \item
    A \textit{jet function} \(J_i( x^2 Q^2R_1^2/\mu^2 )\), describing the subsequent evolution of the energetic parton \(i\) 
    into a jet of ``final-state'' partons on which the PENC is then measured.
    Unlike the hard function, the jet function depends on the parametrization of the observable.
\end{itemize}
In this appendix, we quantify the difference between the jet function for our new PENC (the \(R_1\) jet function) and the traditional PENC (the \(R_L\) jet function). Using theoretical arguments and numerical evidence, we argue that their difference is only relevant at next-to-next-to-leading-logarithmic (NNLL) order.

The resummation of logarithms of $R_1$ in $\Sigma_N$ can be achieved by evaluating the hard function at the scale $\mu_H \sim Q$ and the jet function at the scale $\mu_J \sim Q R_1$, and using the renormalization group equations to evolve them to a common scale $\mu$.
Since the full PENC is independent of \(\mu\), and the hard function is independent of the specific parametrization of the PENC, the renormalization group evolution for the jet function is also parametrization-independent (even if its functional form is not).

Within a jet consisting of two outgoing particles, the PENC we introduce and the traditional PENC are exactly the same, since \(R_1 = R_L\).
Therefore, the \(R_1\) jet function the \(R_L\) jet function are identical at \(\mathcal{O}(\alpha_s)\),
implying that the cross section for the new and traditional PENC will be equal at NLL, as these are single-logarithmic observables.
However, \(R_1\) and \(R_L\) will differ for a jet of three particles or more, corresponding to an ${\cal O}(\alpha_s^2)$ difference between the \(R_1\) and \(R_L\) jet functions, and an NNLL effect in the resummed cross section.

\begin{figure}[h]
    \centering 
    	\includegraphics[width=0.5\textwidth]{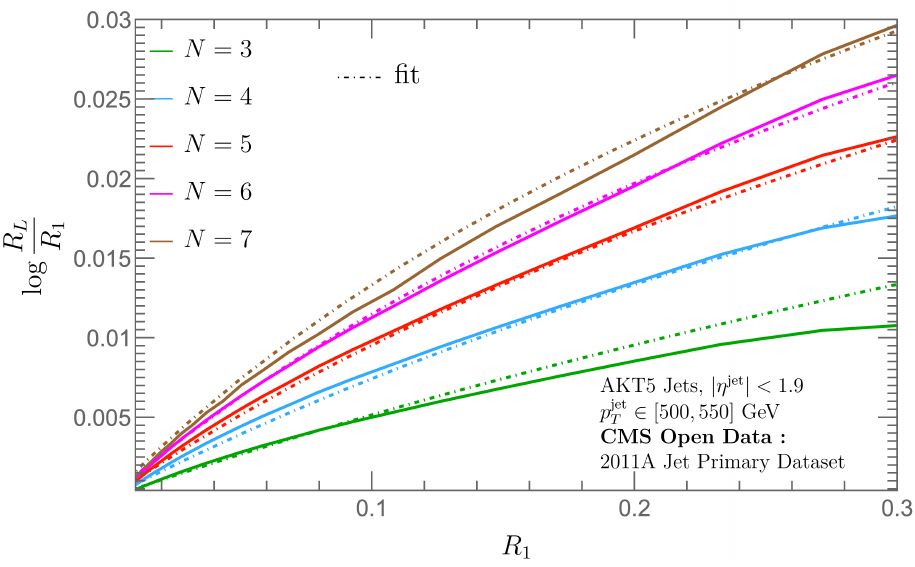}
    \caption{
        Difference between the new parametrization ($R_1$) and the old parametrization ($R_L$) obtained from corresponding quantiles of the cumulative distributions.
        The dashed lines correspond to a fit (of the overall coefficient) of the difference \( \log(R_L/R_1) \) to $\frac{\alpha_s}{\pi} \log(1+R_L\,(N-1))$ for integer values of $N$ from 3 to 7.
    }
	\label{fig:nll_equiv}%
\end{figure}

Though \(R_1\) and \(R_L\) differ in a generic jet, we note that $R_1 \leq R_L \leq 2R_1$ by the triangle inequality.
Therefore, we expect an approximate relationship between the two of the form $R_L \sim (1+c) R_1$ with $0<c<1$.
Furthermore, NLL equivalence of the \(R_1\) and \(R_L\) jet functions implies that  $c \sim {\cal O}(\alpha_s)$.
In particular,
writing the N\(^k\)LL contribution to the traditional PENC as $\sum_{n} d_{n,k}\alpha_s^{n+k}\log^n R_L$ 
and inserting $R_L = (1+c) R_1$ yields: 
\begin{align}
    \label{eq:c_is_alphas}
   \sum_{n,k} 
       d_{n,k} \alpha_s^{n+k}
       \log^n[
           R_1 
           (1 + 
           c)
       ]
   &=
   \sum_{n,k}
      d_{n,k} \alpha_s^{n+k}[\log R_1 +\log(1+c)]^n
  \nn \\
   &=
   \sum_{m,k,k'}
      {m+k' \choose m}
      d_{m+k', k} \alpha_s^{m+k+k'} \log^m R_1 \log^{k'} (1+c)
      \,,
\end{align}
where in the final line we have written \(n = m + k'\).
The terms with \(k' = 0\) give precisely the traditional PENC but with the argument \(R_L\) replaced by \(R_1\).
For \(k' = 1\) this would represent an NLL effect. However, the difference between our new PENC and the traditional PENC starts at NNLL, implying that $\log(1+c) \sim {\cal O}(\alpha_s)$ and thus $c \sim {\cal O}(\alpha_s)$.

Finally, we present additional numerical evidence that \(R_L \sim \left(1+c\right)R_1\) with \(c \sim \mathcal{O}(\alpha_s)\) by using the cumulative distributions for \(R_1\) and \(R_L\) to construct a correspondence between $R_1$ and $R_L$ using quantiles~\cite{Brewer:2018dfs}.
In particular, we numerically invert the equation
\begin{align}
   \label{eq:quantiles}
   \Sigma^\text{new}_{N}(R_1)
   =
   \Sigma^\text{old}_N(
       R_L
       (R_1)
   )
   \, ,
\end{align}
to find a functional form for \(R_L(R_1)\).
In words, \eq{quantiles} selects a function \(R_L(R_1)\) for which our new PENC and the traditional PENC have equivalent cumulative distributions.
\fig{nll_equiv} displays the difference between $\log(R_L(R_1))$ and $\log R_1$, as a function of $R_1$, for $N=$ 3, 4, 5, 6, and 7.
The dashed lines correspond to a fit (of the overall constant) to the form $c \sim \alpha_s/\pi\, \log[1+(N-1) R_L]$.
This provides numerical evidence for our theoretical arguments that the \(R_1\) and \(R_L\) jet functions are equivalent at NLL, and that $R_L = [1 + {\cal O}(\al_s)] R_1$. 

\vspace{15pt}

\section{Samples, Implementation, and Performance}
\label{app:samp_imp_perf}

The studies of this Letter involve both CMS Open Data jet samples from the CMS 2011A Jet Primary Dataset \cite{CERNOpenDataPortal, CMS:JetPrimary2011A} -- also available in MIT Open Data format \cite{Komiske:2019jim, komiske_patrick_2019_3340205} -- and jets in simulated proton-proton collision events generated using \texttt{Pythia 8.310} \cite{Bierlich:2022pfr} with the default Monash tune \cite{Skands:2014pea}.
In \texttt{Pythia 8.310}, we consider jets initiated by QCD interactions, \(W\) bosons, and top quarks, and generate events at hadron level including initial-state radiation, final-state radiation, and multiparton interactions.

Both the CMS 2011A dataset and our simulated \texttt{Pythia} events feature anti-$k_t$ jets \cite{Cacciari:2008gp} with transverse momenta $p_{T,\text{jet}} \in [500, 550]$ GeV and pseudo-rapidity $\vert \eta^{\rm jet} \vert < 1.9$;
however, the CMS dataset involves jets of radius \(R_\text{jet}=0.5\), while our \texttt{Pythia} studies use jets of radius \(R_\text{jet} = 0.8\), which are more appropriate for capturing the substructure of boosted $W$-boson- and top-quark-initiated jets.
Our samples are summarized in Table~\ref{tab:samples}.
Finally, we note that all of the traditional PENCs shown in this work are calculated with the \texttt{EnergyEnergyCorrelators} package~\cite{EEC_github}. 

\begin{table}[ht!]
    \vspace{10pt}
    \centering
    \begin{tabular}{>{\centering\arraybackslash}p{4cm}ccc@{}}
        \toprule
        \textbf{Jet Sample} & \textbf{Center of Mass Energy} & \textbf{AK Radius} 
        \\
        \midrule
        \rowcolor{orange!85!purple!40}
        \textbf{CMS Open Data} & \(7\) TeV & \(R=\,\)0.5 
        \\
        \\
        \midrule
        \texttt{Pythia 8.310} & & & \textbf{Command in \texttt{Pythia}}\\
        \midrule
        \rowcolor{blue!50!green!15}
        \textbf{QCD Jets} & \(14\) TeV & \(R=\,\)0.8 & \texttt{HardQCD:all=on} \\
        \midrule
        \rowcolor{pink!50}
        \textbf{\(\boldsymbol{W}\) Jets} & \(14\) TeV & \(R=\,\)0.8 & \texttt{WeakDoubleBoson:ffbar2WW=on} \\
        \midrule
        \rowcolor{yellow!40}
        \textbf{Top Jets} & \(14\) TeV & \(R=\,\)0.8 & \texttt{Top:gg2ttbar=on}, \texttt{Top:qqbar2ttbar=on} \\
        \bottomrule
    \end{tabular}
    \caption{Jet samples used in this work.
    All jets 
    have transverse momenta $p_{T,\text{jet}} \in [500, 550]$ GeV and pseudo-rapidity $\vert \eta^{\rm jet} \vert < 1.9$.
    The color of each row corresponds to the color in the plots of the associated jet sample.
    \label{tab:samples}
    }

\vspace{45pt}

\begin{minipage}{0.45\textwidth}
\hrule height 1.3pt
\vspace{4pt}
{
    \hypertarget{alg:penc}{ALG I.}\,
    \textbf{Pseudocode} for PENC($R_1$) 
}
\vspace{4pt}
\hrule
\hrule
\vspace{5pt}
\begin{algorithmic}[1]
    \State \textbf{Output:} Normalized 1-dimensional histogram 
    \State \qquad\qquad\,{}
    containing the PENC
    \\
    \State
    \textbf{Initialize} 1D $histogram$ to contain the PENC
    \State \textcolor{codegreen!80!black}{\# Loop over jets in a sample:}
    \ForAll{$i$ \textbf{from} $1$ \textbf{to} $n\_jets$}
        \State $jet \gets$ a jet from the desired sample
        \State \textcolor{codegreen!80!black}{\# Loop over choices for particle $s$ in the jet:}
        \ForAll{particles $s$ in $jet$}
            \State \textcolor{codegreen!80!black}{\# Calculate energy weight}
            \State $z_s \gets $ \(p_{T,\, s}/(\sum_j p_{T,\,j})\)
            \State \textbf{Sort} particles in $jet$ by angle to $s$,
            \State
            \qquad\,\,from \textit{smallest to largest}
            \State \textcolor{codegreen!80!black}{\# Prepare to record total weight within a}
            \State \textcolor{codegreen!80!black}{\# distance \(R_1\) of \(s\), beginning with \(R_1=0\)}
            \State \textbf{Initialize} $sum\_z_1 \gets z_s$
                \State 
                $histogram[0]$
                +=
                $z_s^N$           
            \State \textcolor{codegreen!80!black}{\# Loop over remaining particles $i_1$}
            \ForAll {$i_1$ \textbf{from} $1$ \textbf{to} len(particles)}
                \State $R_1 \gets$ angle between $s$ and particle[$i_1$]
                \State $z_1 \gets $ \(p_{T,\, i_1}/(\sum_j p_{T,\,j})\)
                \State \textcolor{codegreen!80!black}{\# Calculate contribution to $\df\Sigma$, and}
                \State \textcolor{codegreen!80!black}{\# update hist. bin for $R_1$}
                \State 
                $histogram[R_1]$
                +=
                $z_s 
                \,\,\,\,
                \times $
                \\
                \qquad\qquad\qquad\quad
                $\bigl(
                    (
                        sum\_z_1 + z_1
                    )^{N-1}
                    -
                    sum\_z_1^{N-1}
                \bigr)
                $
                \State
                \textcolor{codegreen!80!black}{\# Update sum on weights within \(R_1\)}
                \State
                $sum\_z_1 +\!\!= z_1$
            \EndFor
        \EndFor
    \EndFor
    \\
    \State \textcolor{codegreen!80!black}{\# $histogram$ currently contains $\df\Sigma$, and we want}
    \State \textcolor{codegreen!80!black}{\# PENC$\,=\df \Sigma/\df R_1$}
    \State \textbf{Process} $histogram$ into a probability density by
    \\
    \qquad\qquad\,normalizing it relative to bin widths
    \State \textbf{Write} $histogram$ to output file
\end{algorithmic}
\vspace{5pt}
\hrule height 1.3pt
\end{minipage}
\hfill\vline\hfill
\begin{minipage}{0.45\textwidth}
\hrule height 1.3pt
\vspace{4pt}
{
    \hypertarget{alg:re3c}{ALG II.}\, 
    \textbf{Pseudocode} for RE3C($R_1,R_2,\phi_2$)
}
\vspace{4pt}
\hrule
\hrule
\vspace{5pt}
\begin{algorithmic}[1]
    \State \textbf{Output:} Normalized 3-dimensional histogram 
    \State \qquad\qquad\,{}
    containing the RE3C
    \\
    \State
    \textbf{Initialize} 3D $histogram$ to contain the RE3C
    \State \textcolor{codegreen!80!black}{\# Loop over jets in a sample:}
    \ForAll{$i$ \textbf{from} $1$ \textbf{to} $n\_jets$}
        \State $jet \gets$ a jet from the desired sample
        \State \textcolor{codegreen!80!black}{\# Loop over choices for particle $s$ in the jet:}
        \ForAll{particles $s$ in $jet$}
            \State \textcolor{codegreen!80!black}{\# Calculate energy weight}
            \State $z_s \gets $ \(p_{T,\, s}/(\sum_j p_{T,\,j})\)
            \State \textbf{Sort} particles in $jet$ by angle to $s$,
            \State
            \qquad\,\,from \textit{smallest to largest}
            \State \textcolor{codegreen!80!black}{\# Loop over remaining particles $i_1$ and $i_2$,}
            \ForAll {$i_1$ \textbf{from} $1$ \textbf{to} length(particles)}
                \State $R_1 \gets$ angle between $s$ and particle[$i_1$]
                \State $z_1 \gets $ \(p_{T,\, i_1}/(\sum_j p_{T,\,j})\)
                \State \textcolor{codegreen!80!black}{\# where $i_2$ is \textit{closer} to $s$ than $i_1$}
                \ForAll {$i_2 < i_1$}
                    \State $R_2 \gets$ angle between $s$ and $i_2$
                    \State $z_2 \gets $ \(p_{T,\, i_2}/(\sum_j p_{T,\,j})\)
                    \State $\phi_2 \gets $ angle associated with $i_1$-$s$-$i_2$
                    \State \textcolor{codegreen!80!black}{\# Calculate contribution to $\df\Sigma$,}
                    \State \textcolor{codegreen!80!black}{\# and update associated hist. bin}
                    \State 
                    $histogram[R_1][R_2][\phi_2]$
                    +=
                    $z_s \cdot z_1 \cdot z_2$
                \EndFor
            \EndFor
        \EndFor
    \EndFor
    \\
    \State \textcolor{codegreen!80!black}{\# $histogram$ currently contains $\df\Sigma$, and we want}
    \State \textcolor{codegreen!80!black}{\# RE3C$\,=\df \Sigma/\df R_1 \df R_2 \df \phi_2$}
    \State \textbf{Process} $histogram$ into a probability density by
    \\
    \qquad\qquad\,normalizing it relative to bin widths
    \State \textbf{Write} $histogram$ to output file
\end{algorithmic}
\vspace{8pt}
\hrule height 1.3pt
\end{minipage}
\end{table}

\vspace{10pt}

We also provide pseudocode of our new algorithms for calculating PENCs and RE3Cs in Algs.~\hyperlink{alg:penc}{I} and  \hyperlink{alg:re3c}{II}, respectively.
Alg.~\hyperlink{alg:penc}{I} exemplifies how the parametrization we introduce for the PENC (i.e.~of angles with respect to a specific particle $s$) simplifies the $N$-dependence of our new PENC, dramatically speeding up the associated computation.
In particular, the algorithm has identical runtime behavior for any value of \(N\), and \emph{there are no additional computational complications for large or non-integer \(N\).}
For completeness, we also present pseudocode for the RE3C in Alg. \hyperlink{alg:re3c}{II}, which features similar computational simplicity and can be easily generalized to a double-projected ENC as discussed in the main text.
For a more complete and technical description of each computation, including contributions in which the same particle enters multiple times in the correlator (often called contact terms), please consult our full implementation on GitHub at \texttt{ResolvedEnergyCorrelators}~\cite{github:RENC}.

\begin{figure}[t]
    \centering 
    \subfloat[]{
        \includegraphics[width=.45\textwidth]{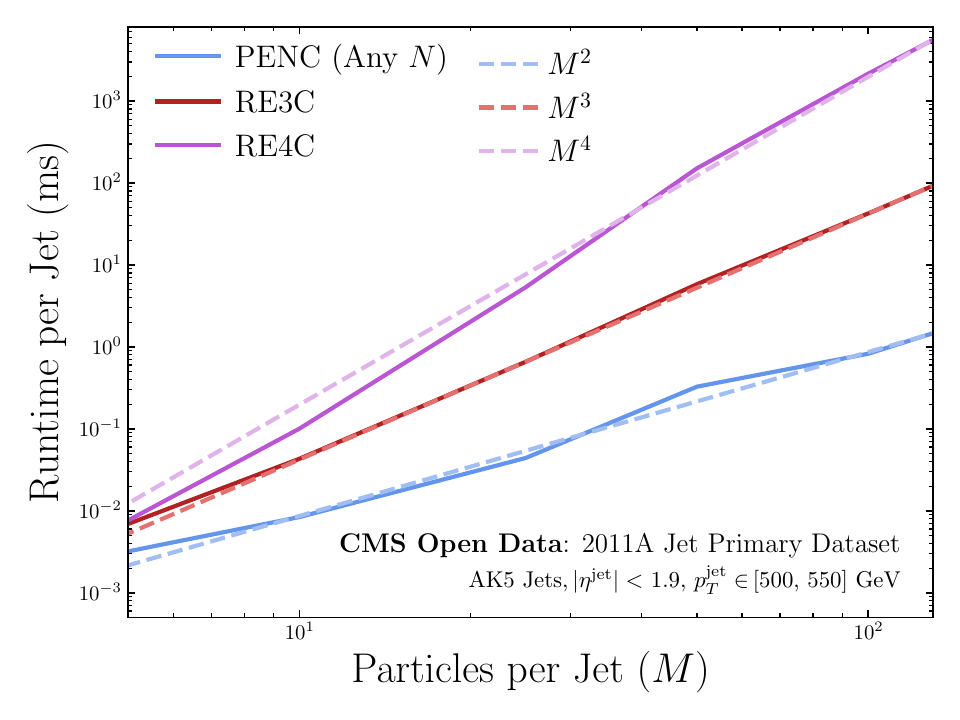}
        \label{fig:runtime_scaling}
    }
    $\qquad$
    \subfloat[]{
        \includegraphics[width=.45\textwidth]{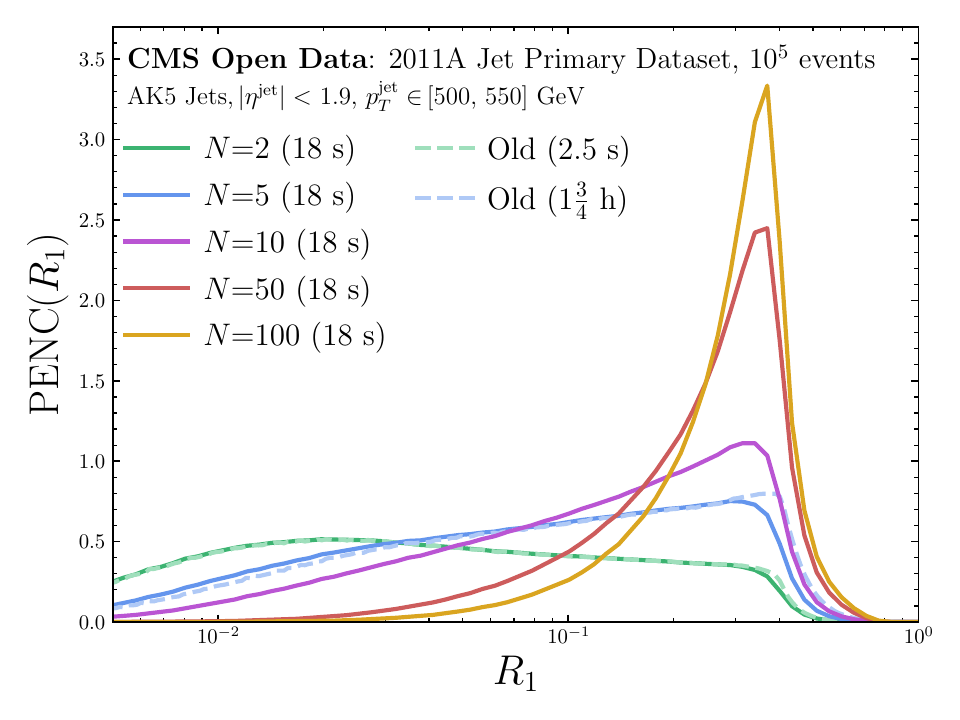}    
        \label{fig:runtime_largeN}
    }
    \caption{
        Plots emphasizing the computational performance of the energy correlators we introduce in this work.
        \textbf{(a)}
        Runtime of the PENC and RENC code of Ref.~\cite{github:RENC} on CMS Open Data as a function of the number of particles in a jet, together with a polynomial fit to guide the eye.
        \textbf{(b)}
        PENCs for $N=$ 2, 5, 10, 50, and 100, each computed using on \(10^5\) CMS Open Data jets in less than thirty seconds.
        The solid lines show the PENCs introduced in this work, and the dashed lines show the traditional PENCs.
        Even for $N=5$, the traditional computation takes $6266$ seconds (close to 2 hours) to run on the same CMS Open Data samples, and the computation time grows exponentially as $N$ increases.
        The large values of $N$, which are computationally inaccessible to the traditional PENC even for $N=10$, are chosen to stress that the PENCs we introduce in this work make quick work of previously unimaginable computations. 
    }
	\label{fig:runtime}%
\end{figure}

Finally, in \fig{runtime}, we display the computational runtime of our code for the evaluation of PENCs and RENCs on jet samples from CMS Open Data.
\fig{runtime_scaling} shows how the scaling time for each nearly follows an \(M^r\) scaling, where \(r\) indicates the number of resolved emissions.
Furthermore, the independence of the runtime on the value of \(N\) allows us to compute the PENC for enormous, and previously unimaginable, values of \(N\);
some examples up to \(N = 100\) are shown in \fig{runtime_largeN}.
We highlight that our implementation takes similar runtime for all values of $N$.
Therefore, we expect the PENC we introduce in this work to significantly benefit ongoing studies which apply energy correlators in jets produced in heavy-ion collisions for revealing the emergent scales of the quark-gluon plasma in these environments. 

\section{Visualizing CMS Open Data with Resolved Correlators}
\label{app:opendata}

In this appendix, we provide additional polar heat maps, or \textit{bullseye} visualizations, of RE3Cs and RE4Cs evaluated on CMS Open Data, which yield an intuitive view into the internal structure of jets enabled by our new parametrization.
We expect that similar visualizations will benefit current and future studies of jet substructure at the LHC.

\begin{figure}[p]
    \vspace{-20pt}
    \centering 
    \subfloat[]{
    	\includegraphics[width=0.32\textwidth]{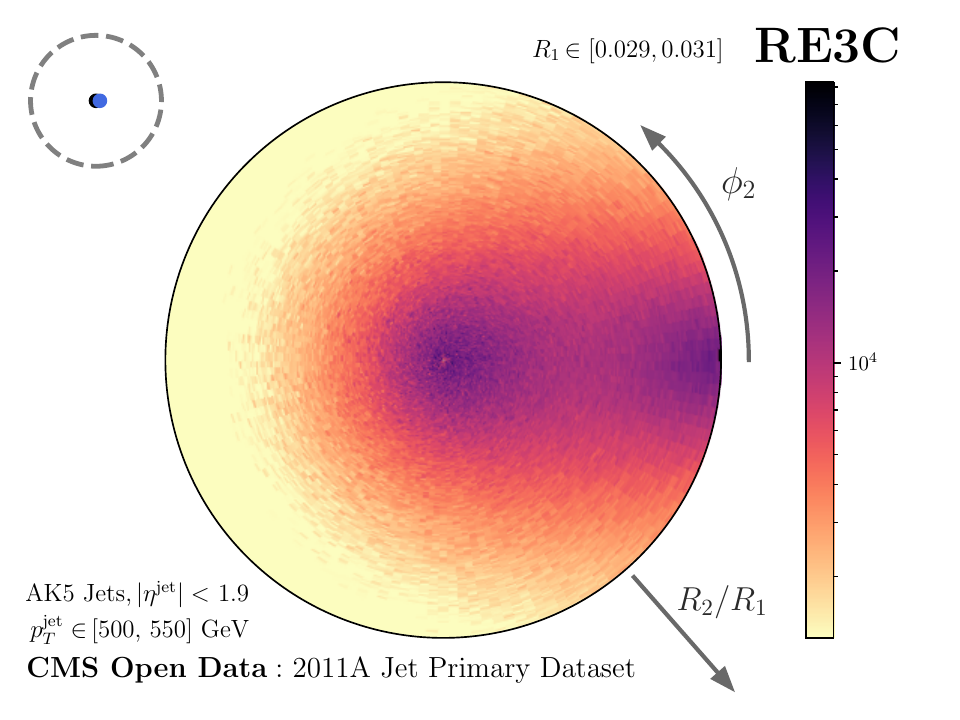}
     }
    \subfloat[]{
    	\includegraphics[width=0.32\textwidth]{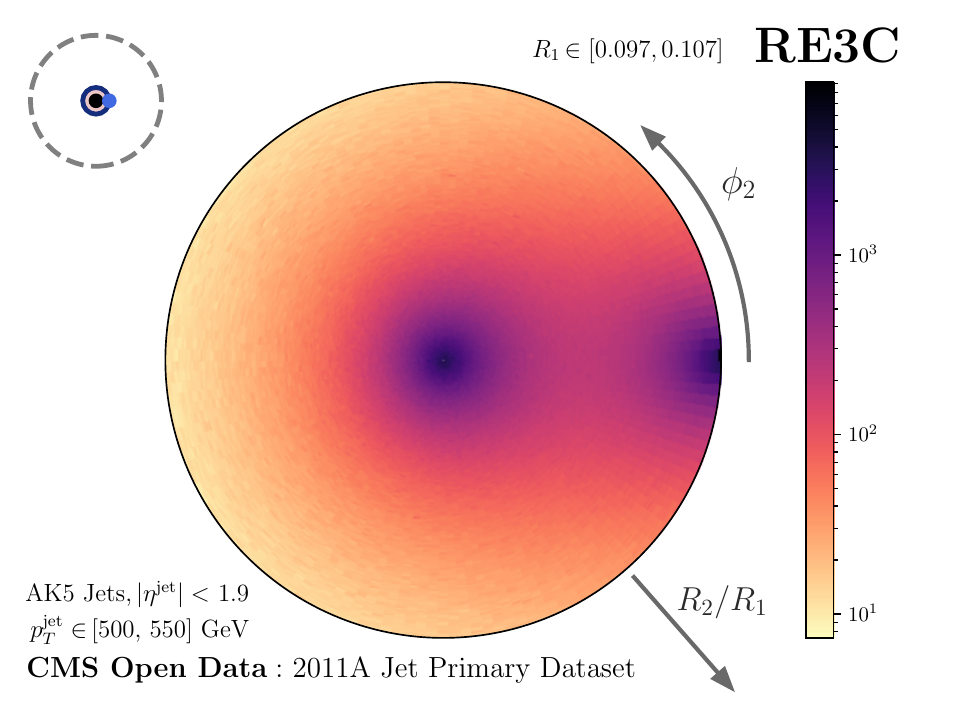}
     }
    \subfloat[]{
    	\includegraphics[width=0.32\textwidth]{figures/opendata/od_3particle_bullseye_0.286592.pdf}
     }
    \caption{
        Additional polar heat maps for RE3Cs evaluated on CMS Open Data.
        There are enhanced correlations when \(R_2\sim 0\), corresponding to the collinear enhancement 
        as \(i_2\) approaches \(s\), and when \(\phi_2 \sim 0\), \(R_2 \sim R_1\), corresponding to \(i_2 \to\) \(i_1\).
        The \(R_1\) bin for each plot is indicated by the radius of the filled circle in the top-left inset of each plot.
    }
    \label{fig:cms_re3cs}
\end{figure}

\begin{figure}[p]
    \centering 
    \subfloat[]{
    	\includegraphics[width=0.32\textwidth]{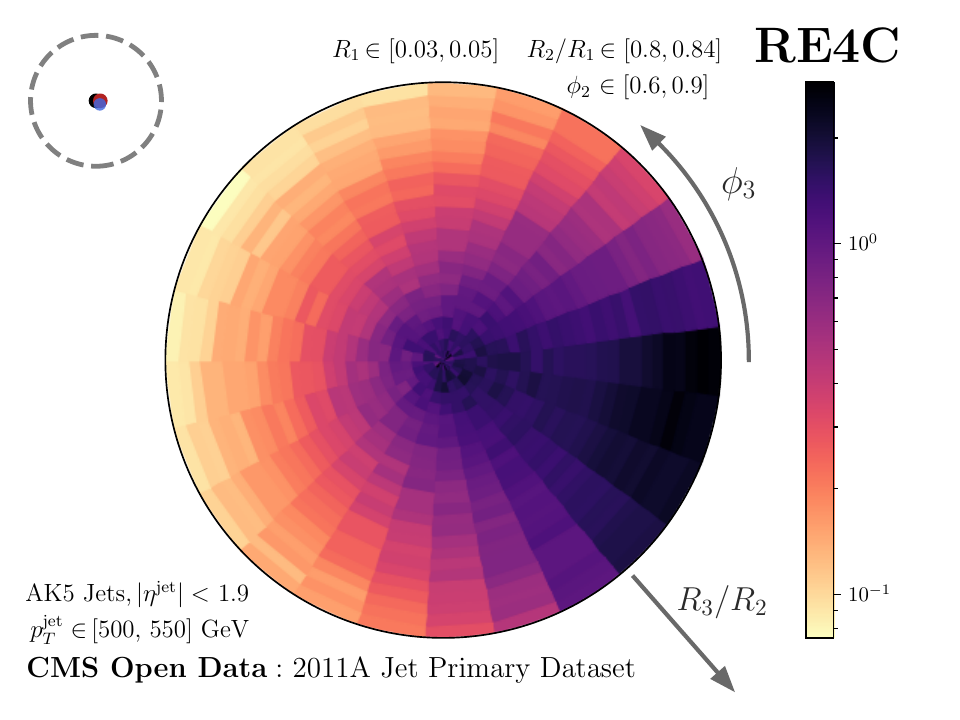}
    }
    \subfloat[]{
    	\includegraphics[width=0.32\textwidth]{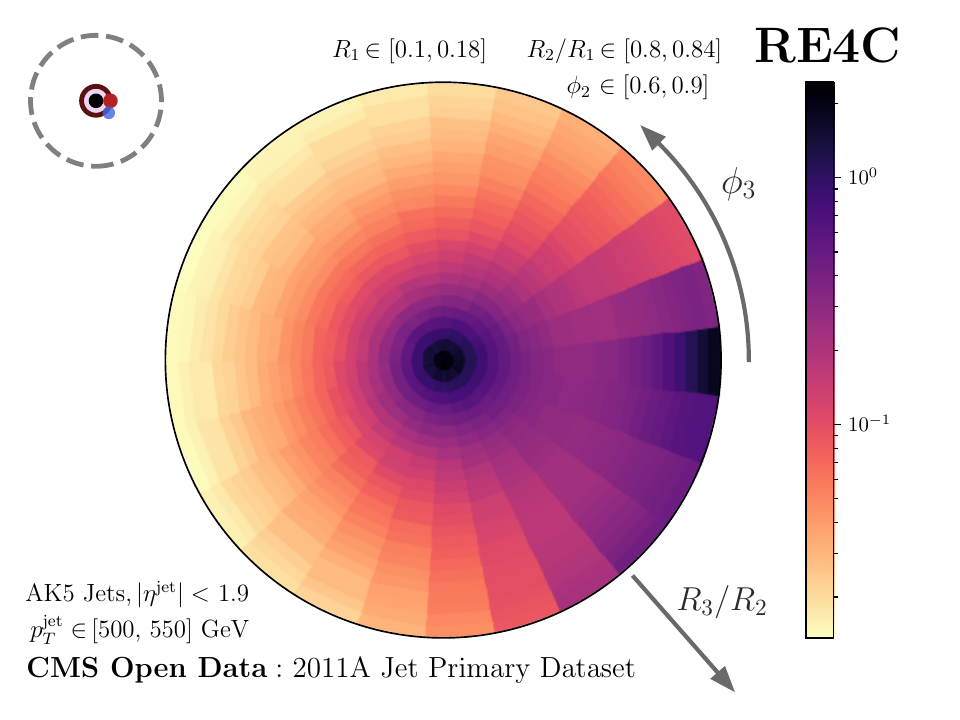}
    }
    \subfloat[]{
    	\includegraphics[width=0.32\textwidth]{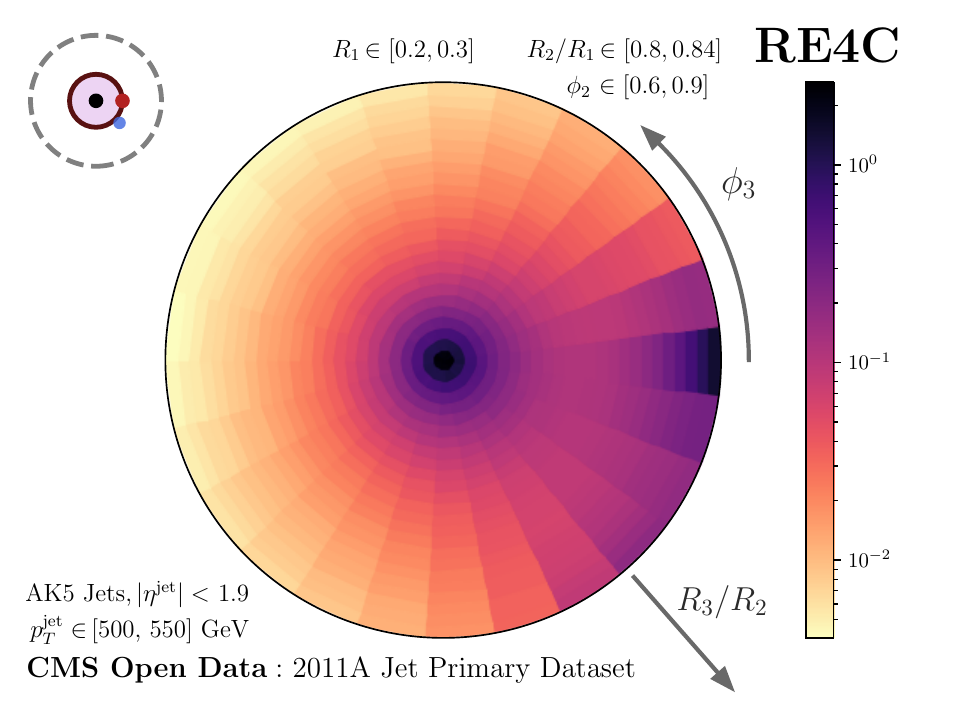}
     }
    \\
    \vspace{-8pt}
     \subfloat[]{
    	\includegraphics[width=0.32\textwidth]{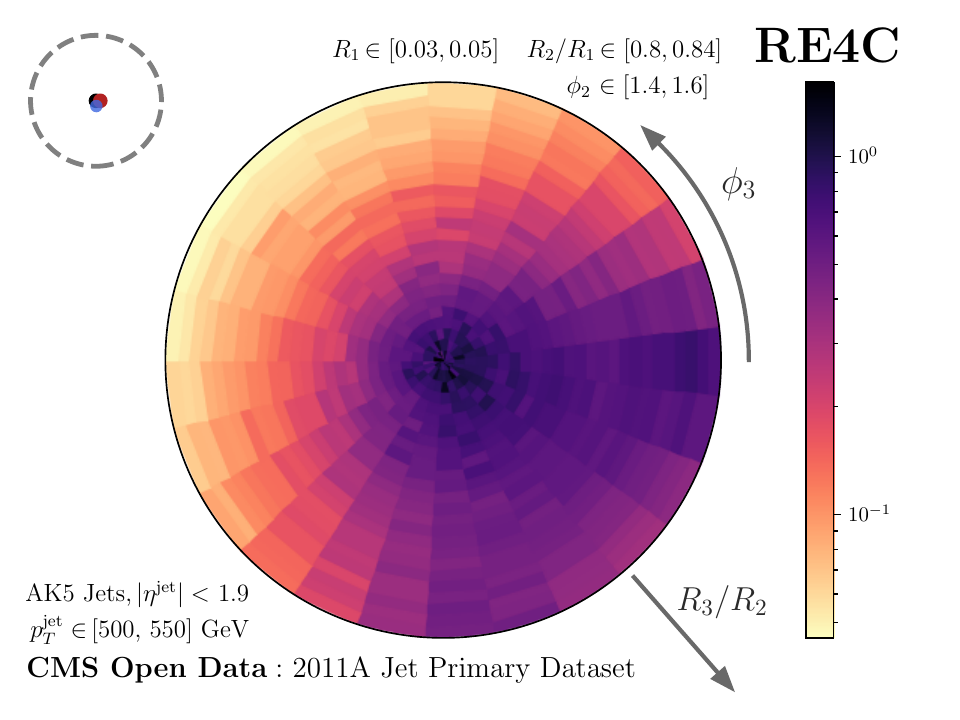}
    }
    \subfloat[]{
    	\includegraphics[width=0.32\textwidth]{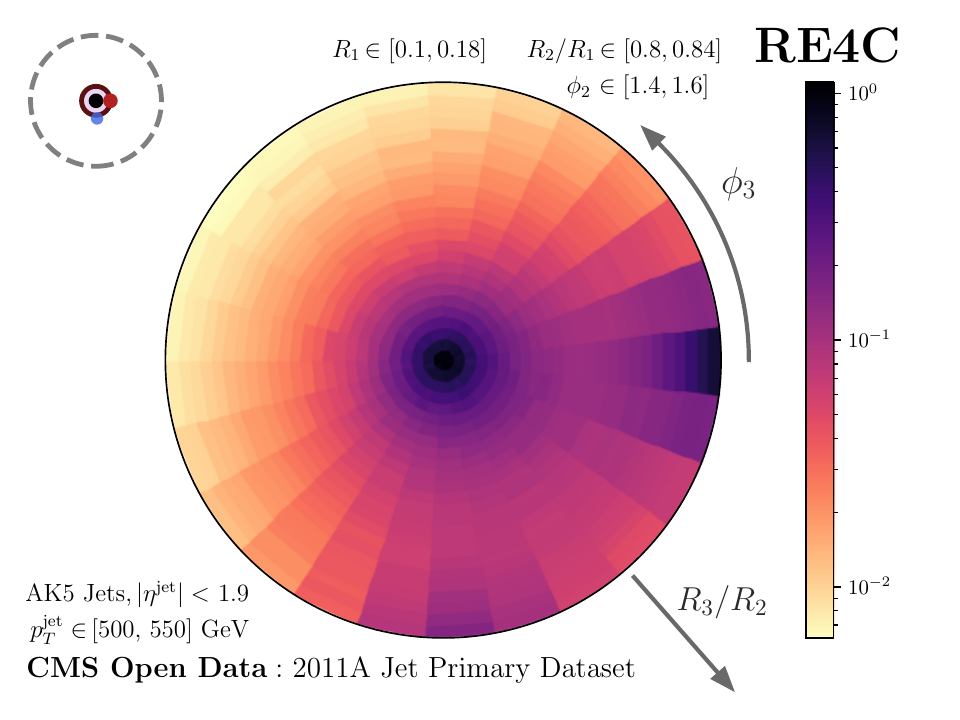}
    }
    \subfloat[]{
    	\includegraphics[width=0.32\textwidth]{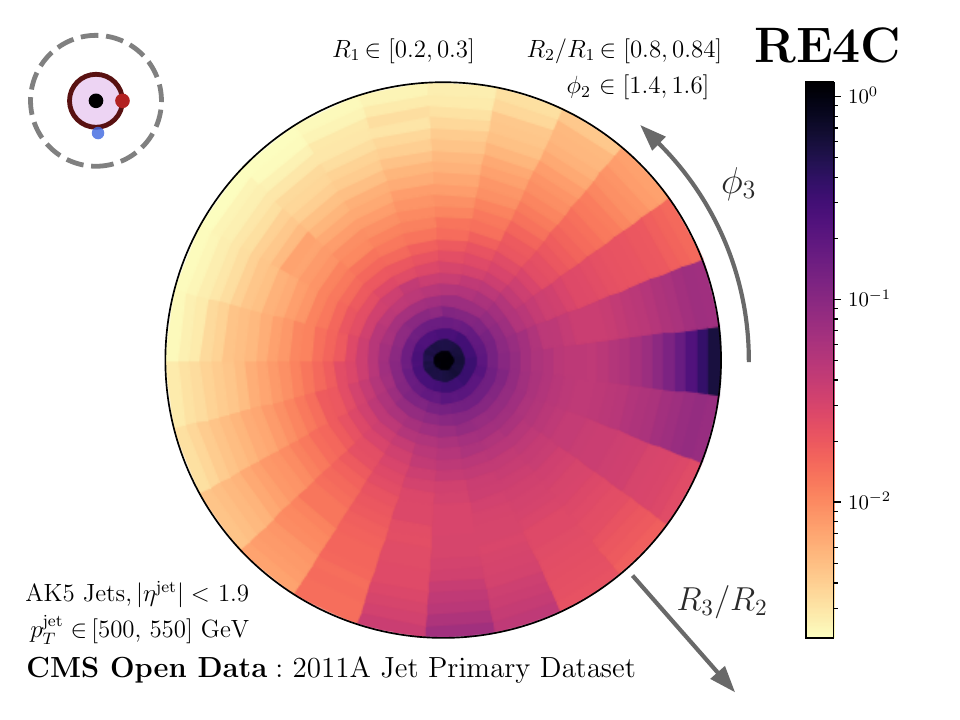}
    }
    \\
    \vspace{-8pt}
     \subfloat[]{
    	\includegraphics[width=0.32\textwidth]{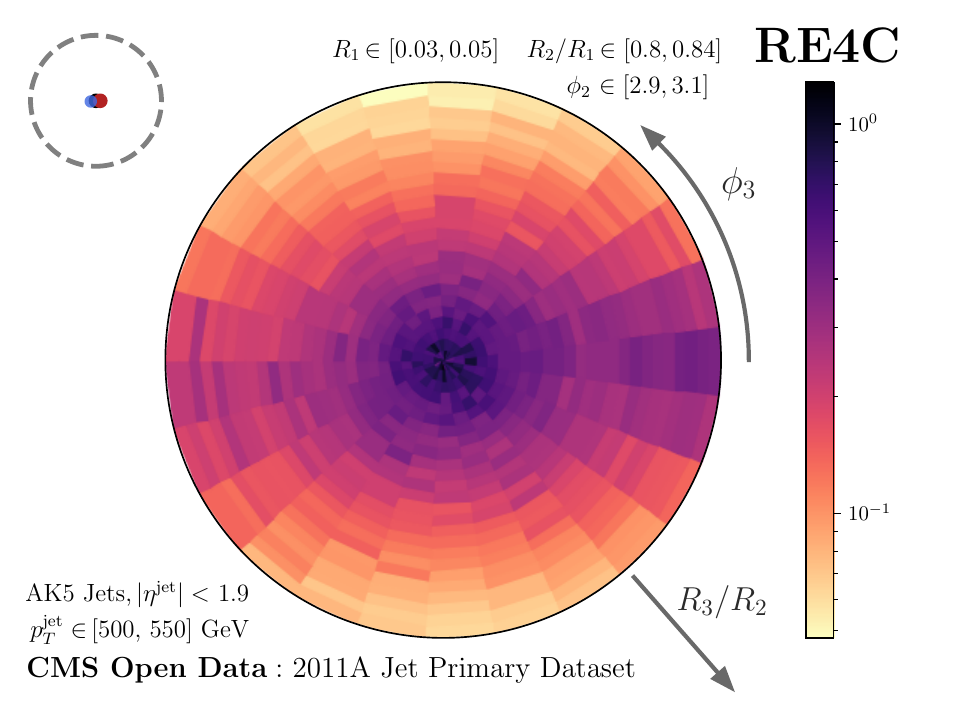}
    }
    \subfloat[]{
    	\includegraphics[width=0.32\textwidth]{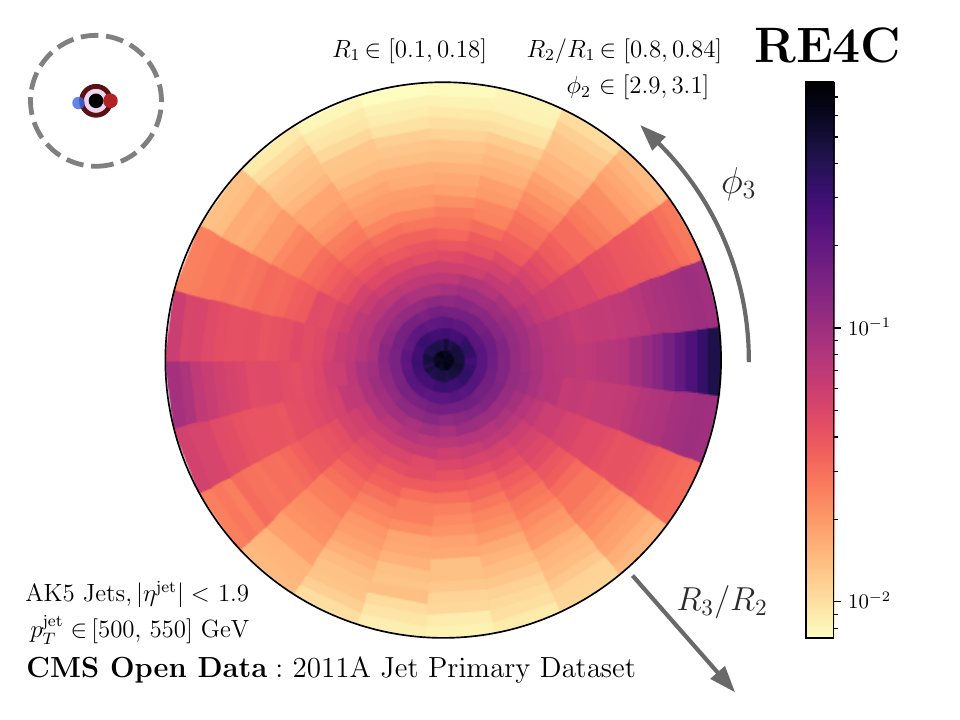}
    }
    \subfloat[]{
    	\includegraphics[width=0.32\textwidth]{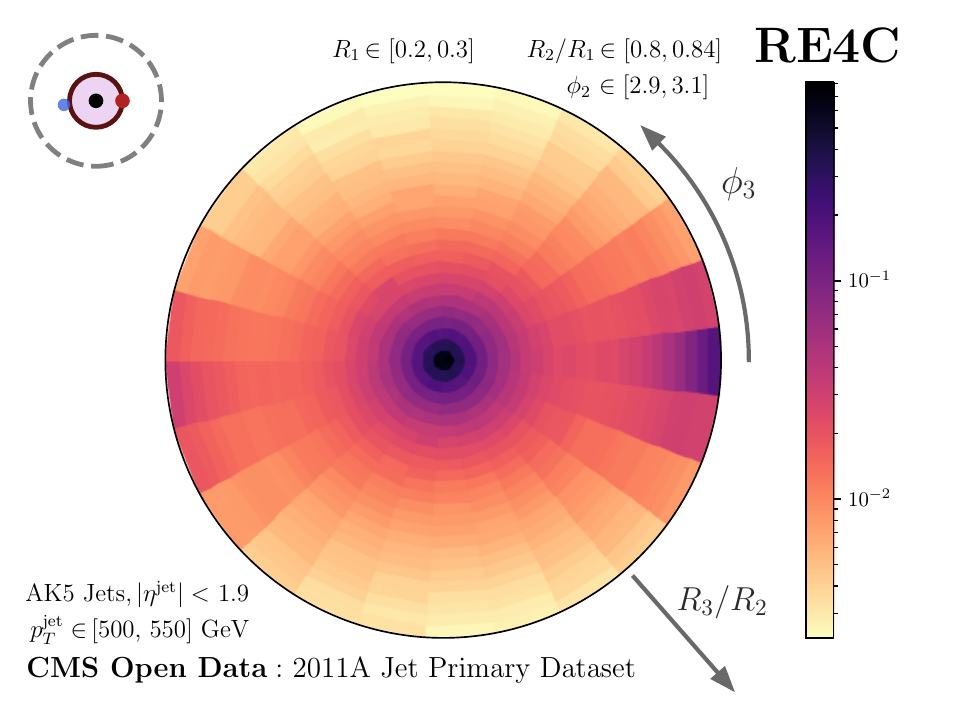}
    }
    \caption{
        Polar heat maps for RE4Cs evaluated on CMS Open Data, with \(\phi_2\) centered near $45^\circ$ (top row), $90^\circ$ (middle row) and $180^\circ$ (bottom row).
        We highlight that, relative to the RE3C presented in \fig{cms_re3cs}, there are enhanced correlations when \(\phi_3 \sim -\phi_2\), corresponding to the collinear enhancement of radiation as \(i_3\) approaches \(i_1\).
    }
    \label{fig:cms_re4cs}
\end{figure}

The bullseyes for the RE3C, shown in \fig{cms_re3cs}, have a radial variable corresponding to the ratio \(R_2/R_1 < 1\), with a polar angle corresponding to \(\phi_2\).
Similarly, the bullseyes for the RE4C in \fig{cms_re4cs} use the radial variable \(R_3/R_2 < 1\) and the corresponding polar variable \(\phi_3\).
The bullseyes are normalized to a radial measure, such that the bullseye plots provide a faithful representation of each correlator;
for the RE3C, the bullseye is normalized to unity against the measure \(\df\log R_1 \times R_2 \df R_2 \df \phi_2\), while the bullseyes for the RE4C are normalized to unity against \(\df \log R_1 \df (R_2/R_1) \df \phi_2  \times R_3 \df R_3 \df \phi_3\)

RE3Cs are shown in \fig{cms_re3cs} for several \(R_1\) bins, and RE4Cs are shown in \fig{cms_re4cs} for several \(R_1\) and \(\phi_2\) bins with a fixed \(R_2/R_1\) bin.
We note that the RE3C distributions are quite similar across different \(R_1\) bins.
On the other hand, for \(R_2\) near \(R_1\), the RE4C distributions exhibit distinct patterns of radiation depending on the value of \(\phi_2\).
In particular, \fig{cms_re4cs} prominently visualizes the collinearly enhanced correlations in the RE4C when \(\phi_3\) is near \(-\phi_2\) and \(R_3\) is near \(R_1\), i.e.~when particle \(i_3\) approaches particle \(i_1\).
For small \(R_2\), these correlations diminish, and the bullseye visualizations for the RE4C look very similar to that of the RE3C, reflecting the near-scale-invariance of QCD.


\section{Process Dependence: Pythia Samples}
\label{app:pythia}

\begin{figure}[t]
    \centering 
    \subfloat[]{
    	\includegraphics[width=0.3\textwidth]{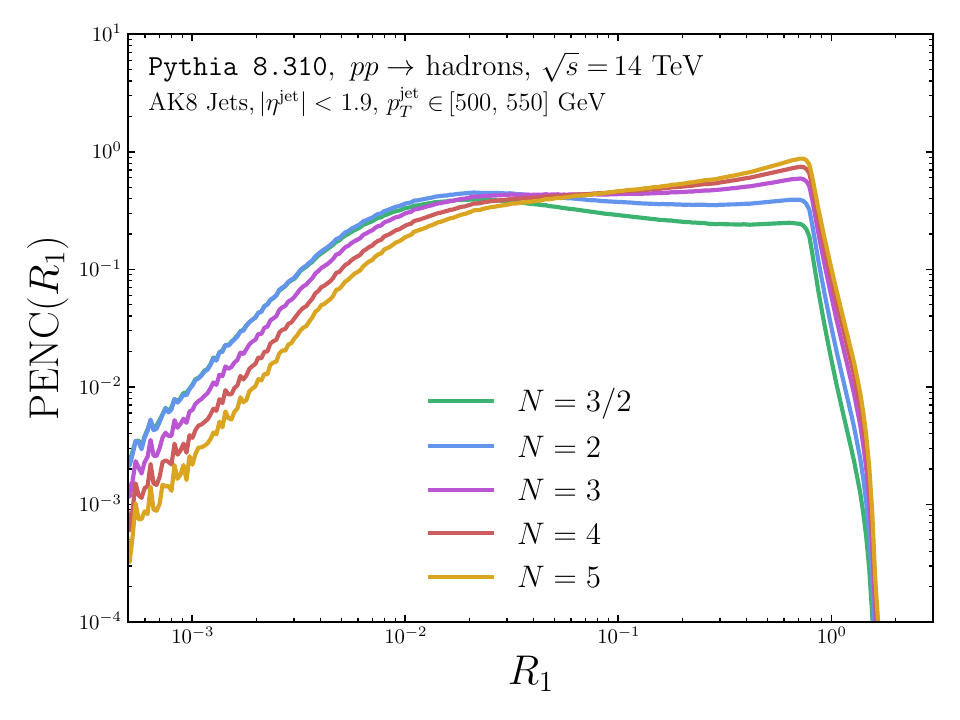}
    }
    $\quad$
    \subfloat[]{
    	\includegraphics[width=0.3\textwidth]{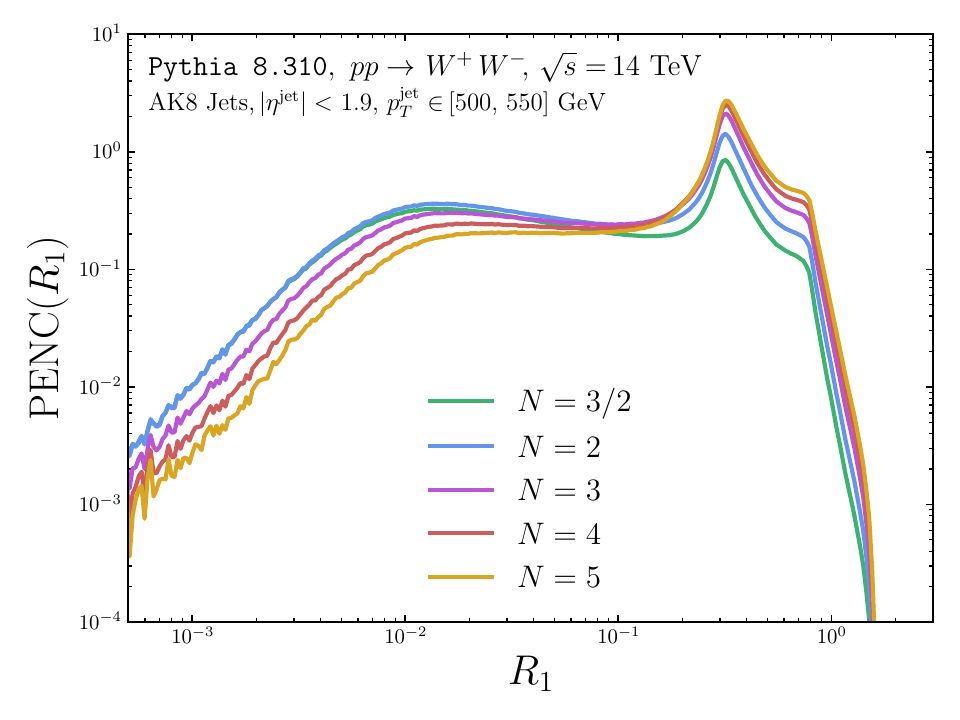}
    }
    $\quad$
    \subfloat[]{
    	\includegraphics[width=0.3\textwidth]{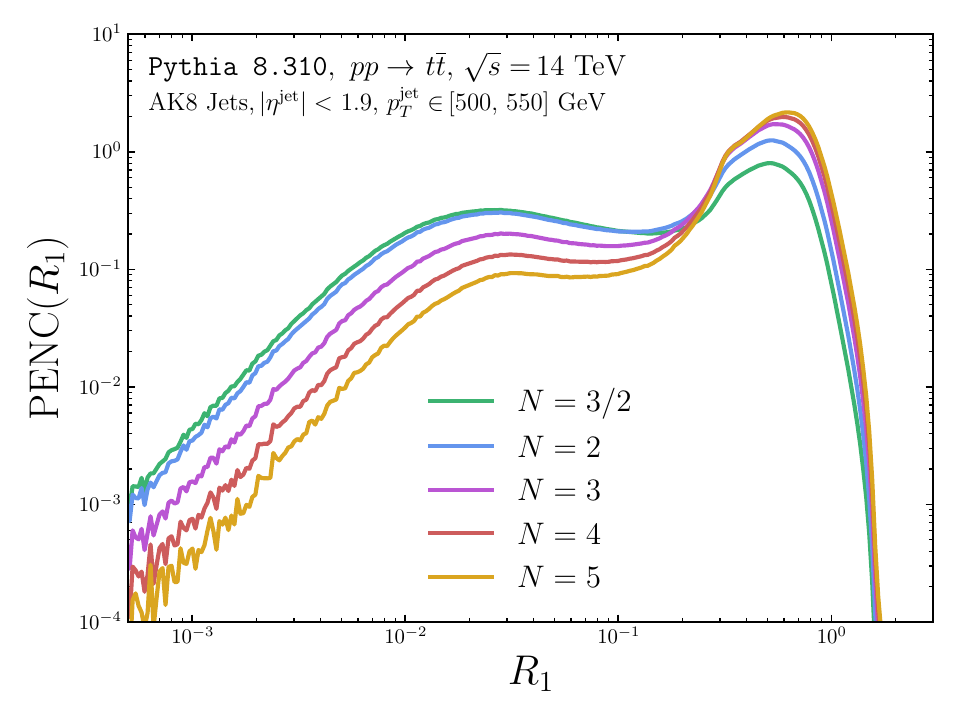}
    }
    \caption{
        PENCs for {\textbf{(a)}} QCD-, {\textbf{(b)}} $W$- and {\textbf{(c)}} top-quark-initiated jet samples generated with \texttt{Pythia 8.310}.
        Jets are clustered using the anti-$k_t$ algorithm with a radius parameter of $R=0.8$, and have transverse momenta in the range $p_T^{\rm jet} \in [500,550] \,\text{GeV}$ and psuedo-rapidities in the range $\vert \eta^{\rm jet}\vert < 1.9$.
        PENCs for $W$- and top-quark-initiated jets feature enhanced correlations at scales correlated with the $W$-boson and top-quark masses.
    }
	\label{fig:pythia_pencs}%
\end{figure}

This appendix provides visualizations for energy correlators evaluated on Monte Carlo samples of QCD-, \(W\)-, and top-initiated jets generated in \texttt{Pythia 8.310}, as detailed in Table~\ref{tab:samples}, which emphasize both the broad range of applications and the visually intuitive representations of jet substructure that are achieved by our new parametrization for RE3Cs.
We begin by showing PENCs evaluated on each jet sample in \fig{pythia_pencs}, which provide simple, albeit less fetching, visualizations which clearly distinguish each sample of jets.

\begin{figure}
    \centering 
    \subfloat[]{
    	\includegraphics[width=0.34\textwidth]{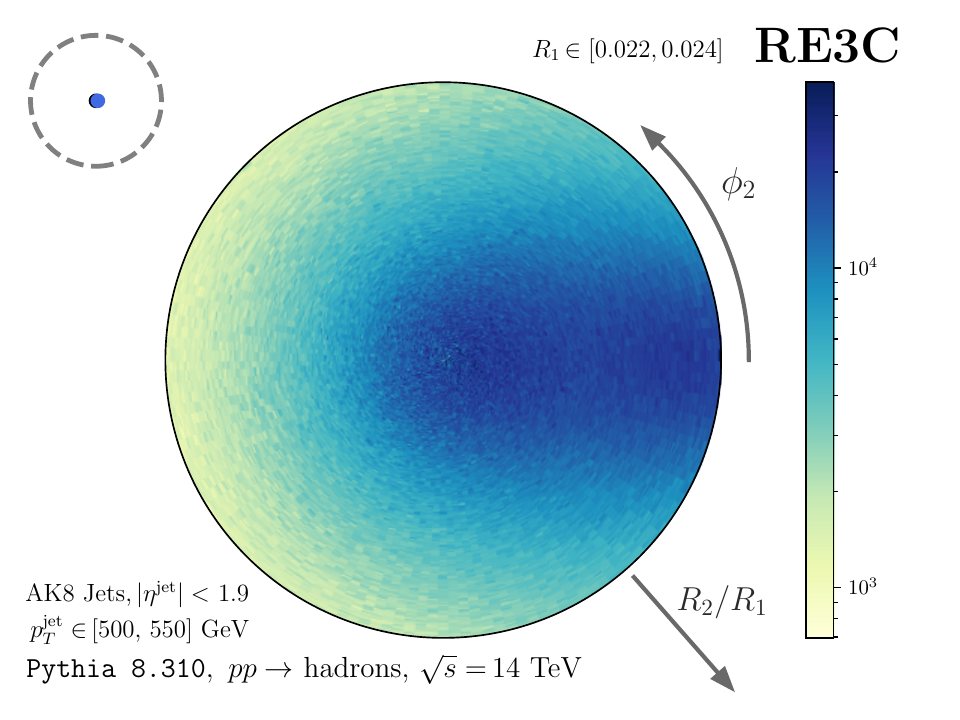}
    }
    \subfloat[]{
    	\includegraphics[width=0.34\textwidth]{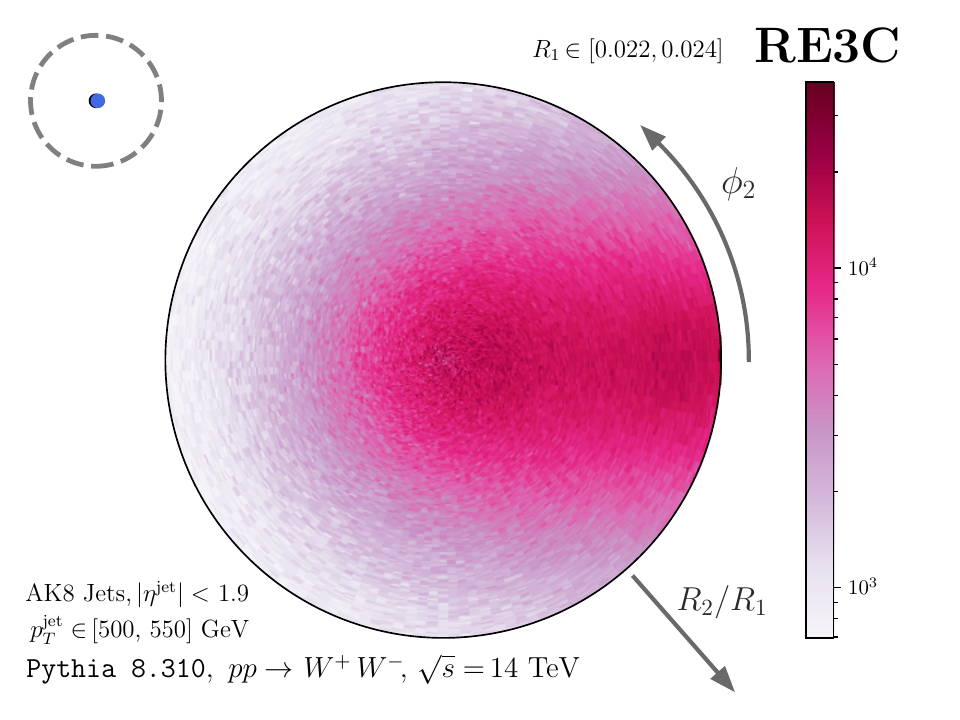}
    }
    \subfloat[]{
    	\includegraphics[width=0.34\textwidth]{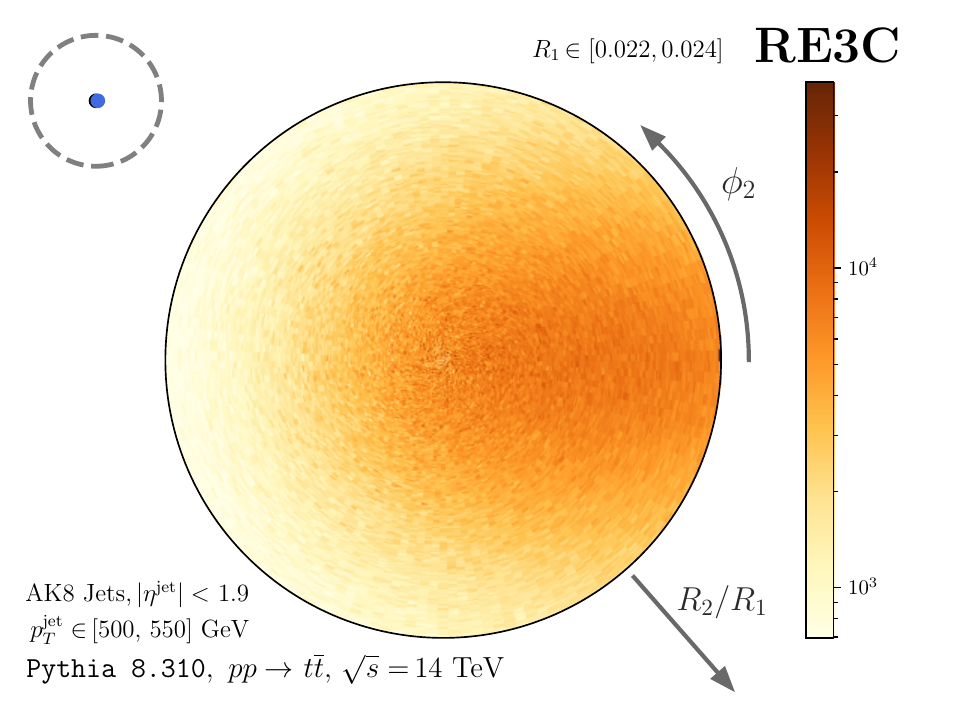}
    }
    \\
    \subfloat[]{
    	\includegraphics[width=0.34\textwidth]{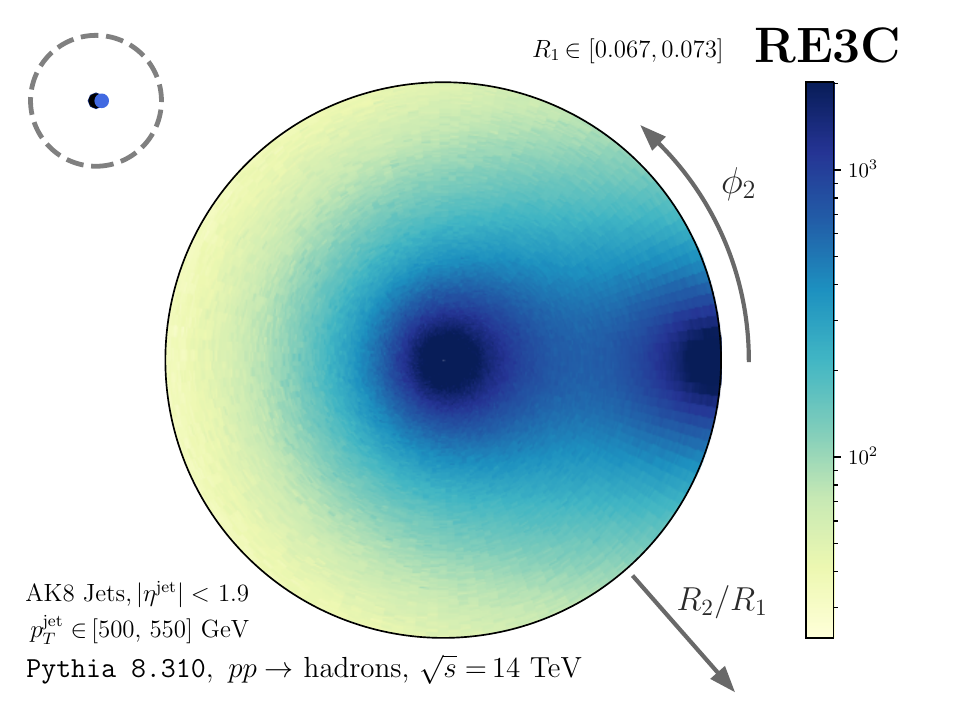}
    }
    \subfloat[]{
    	\includegraphics[width=0.34\textwidth]{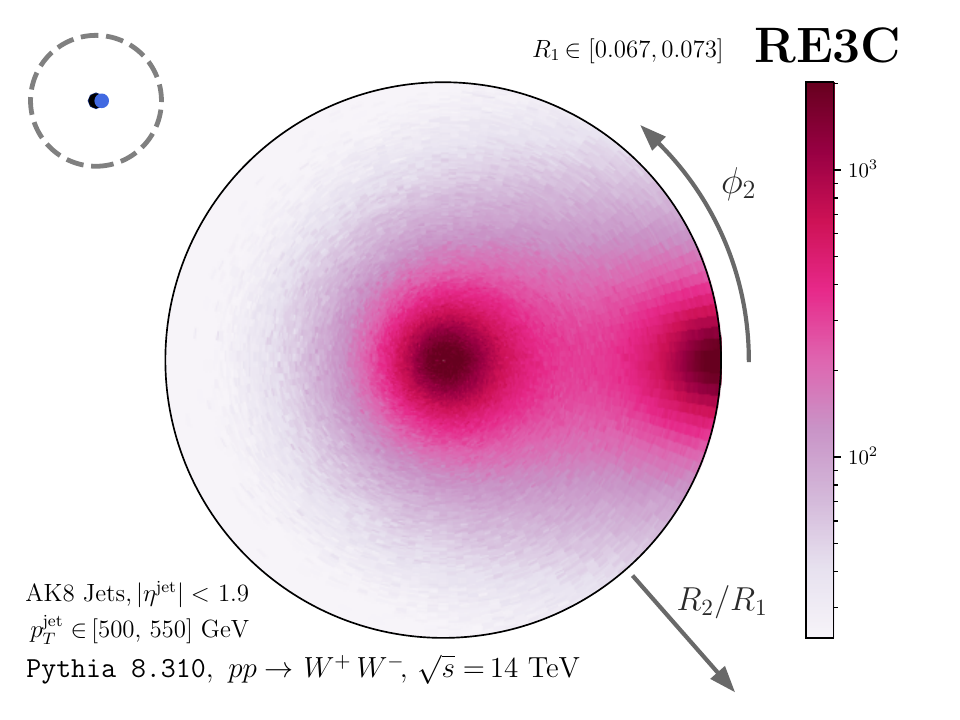}
    }
    \subfloat[]{
    	\includegraphics[width=0.34\textwidth]{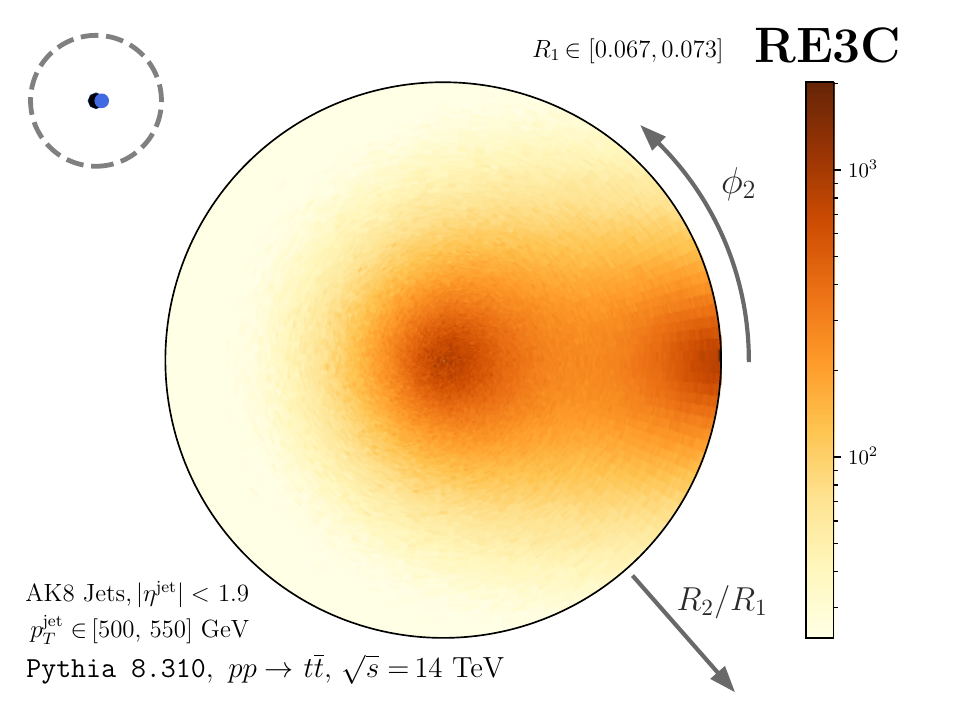}
    }
    \\
    \subfloat[]{
    	\includegraphics[width=0.34\textwidth]{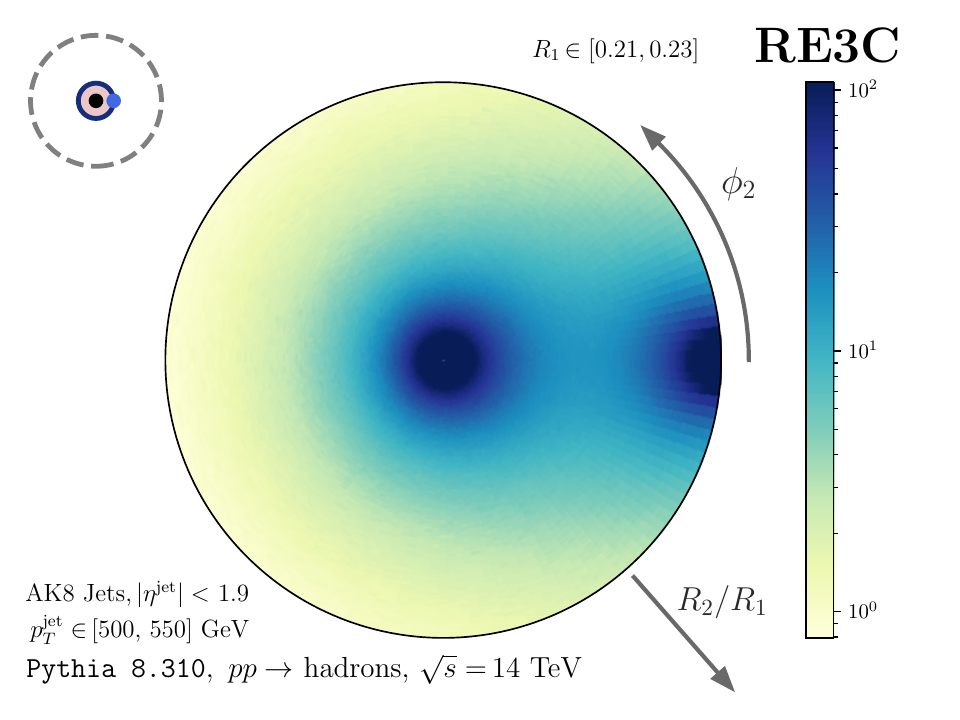}
    }
    \subfloat[]{
    	\includegraphics[width=0.34\textwidth]{figures/supplemental/3particle/w0.216087_3particle_bullseye.pdf}
        \label{fig:w_medium_re3c}
    }
    \subfloat[]{
    	\includegraphics[width=0.34\textwidth]{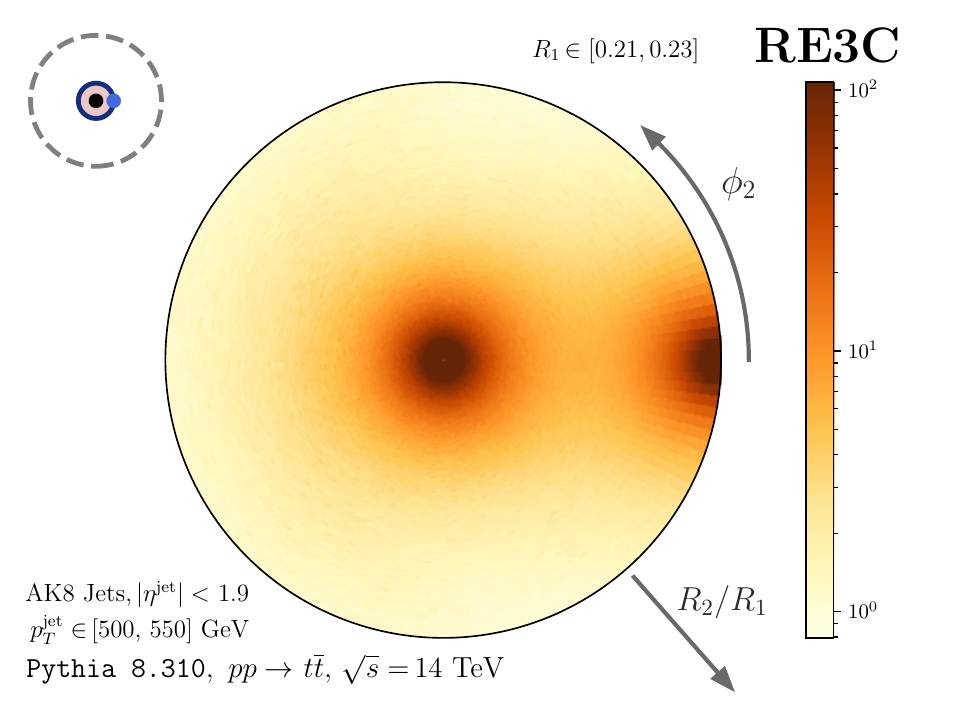}
    }
    \\
    \subfloat[]{
    	\includegraphics[width=0.34\textwidth]{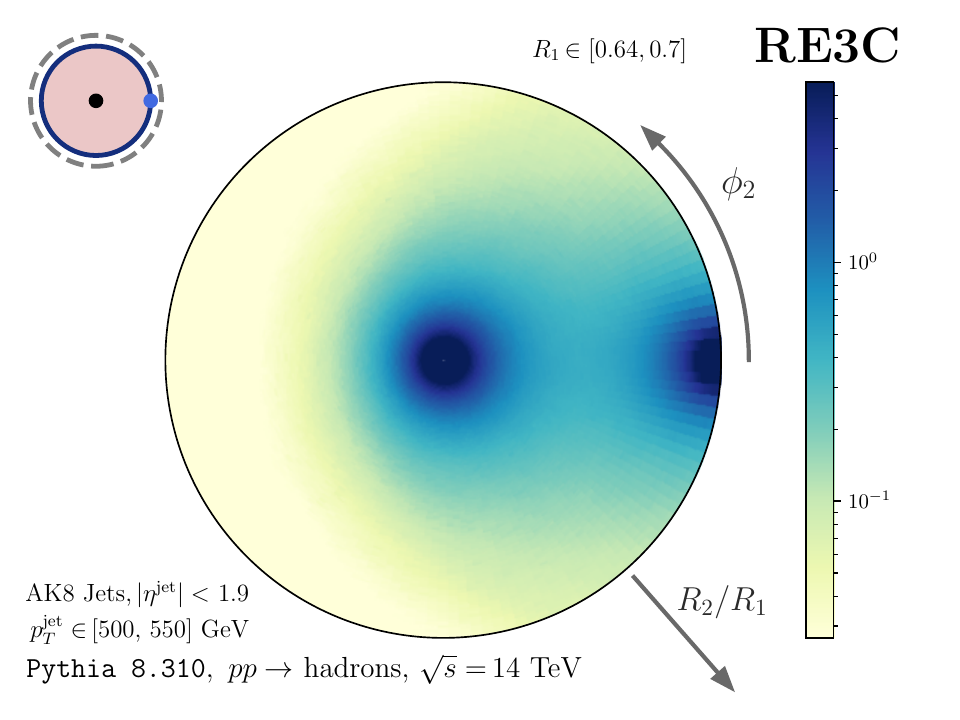}
    }
    \subfloat[]{
    	\includegraphics[width=0.34\textwidth]{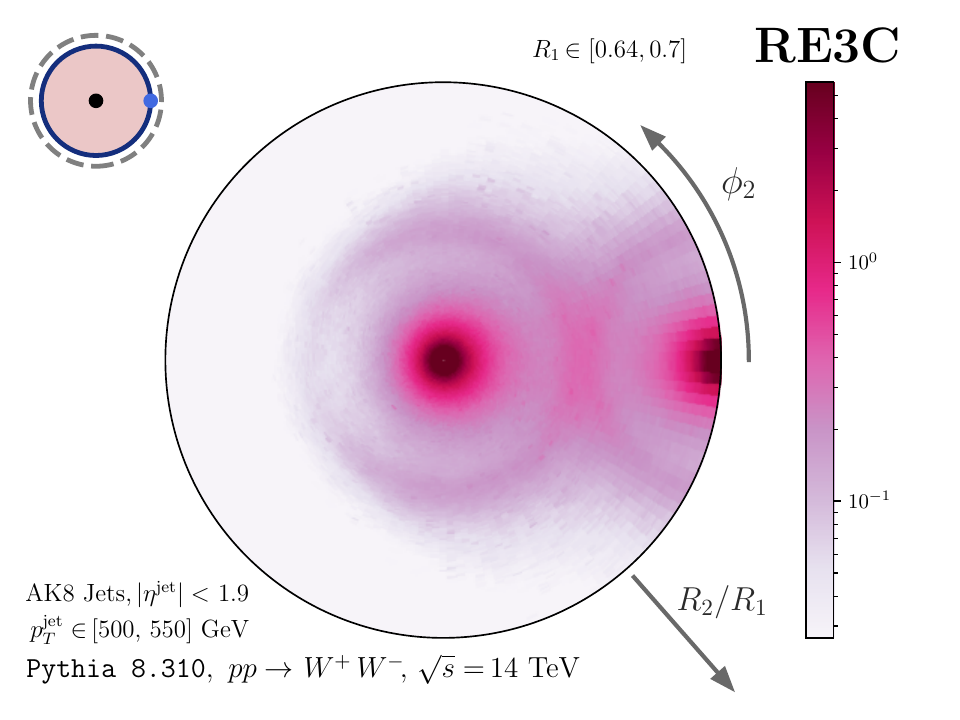}
        \label{fig:w_large_re3c}
    }
    \subfloat[]{
    	\includegraphics[width=0.34\textwidth]{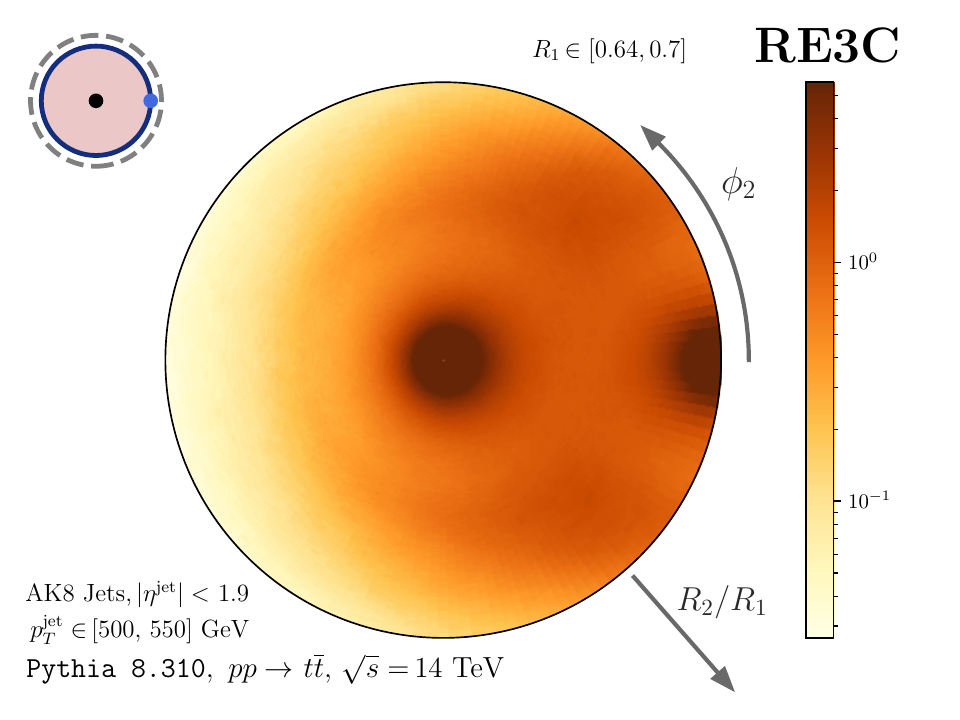}
    }
    \caption{
        Polar heat maps of the RE3C we introduce in this work for (first column) QCD-, (second column) \(W\)-, and (third column) top-quark-initiated jets.
        The radial direction in each plot is the ratio \(R_1/R_2\), and the polar angle of each plot indicates the angle \(\phi_2\).
        Our RE3Cs provide a clear visual representation of the distinct patterns of radiation in each jet sample at different scales (set by \(R_1\)).
    }
	\label{fig:pythia_re3cs}%
\end{figure}

The visually distinct features of each jet sample are even more evident as more particles are resolved, i.e.~by the RE3C and RE4C.
In \fig{pythia_re3cs}, we present bullseye visualizations of the RE3C for each jet sample in several \(R_1\) bins;
as in the plots using CMS Open Data in the previous appendix, the radial variable and polar angle for each RE3C bullseye are the ratio \(R_2/R_1\) and the angle \(\phi_2\), respectively.
We note that the rings in the polar heat maps for the RE3C are correlated with the mass of the \(W\) boson, in that they appear at the angular scale \(R_2 \simeq 0.3 \sim 2 m_W / p_T^\text{jet}\).

\begin{figure}
    \centering 
    \subfloat[]{
    	\includegraphics[width=0.34\textwidth]{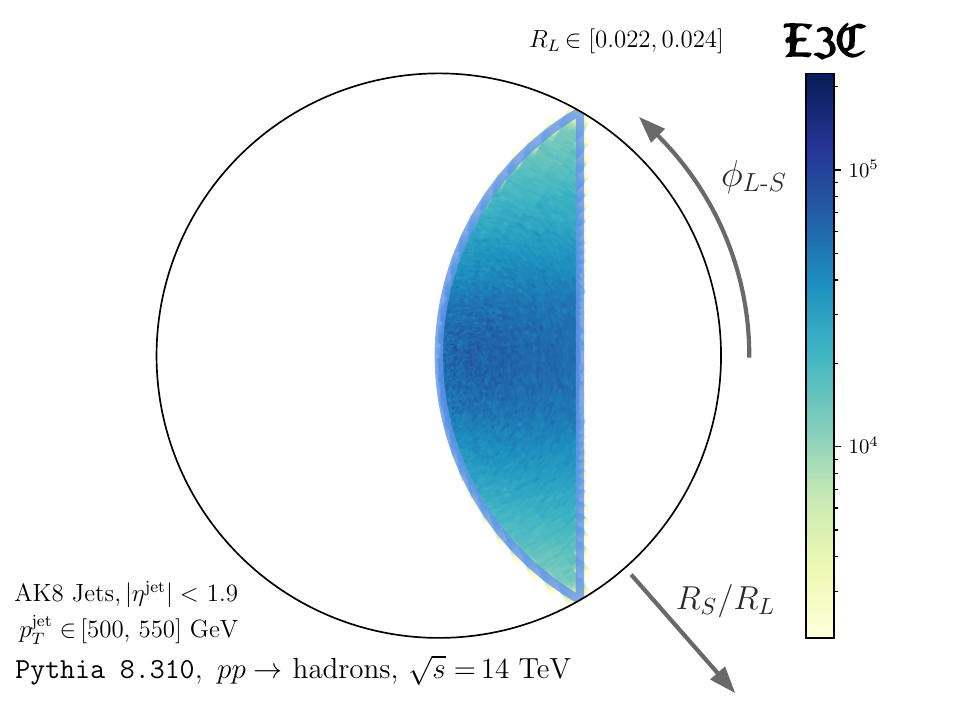}
    }
    \subfloat[]{
    	\includegraphics[width=0.34\textwidth]{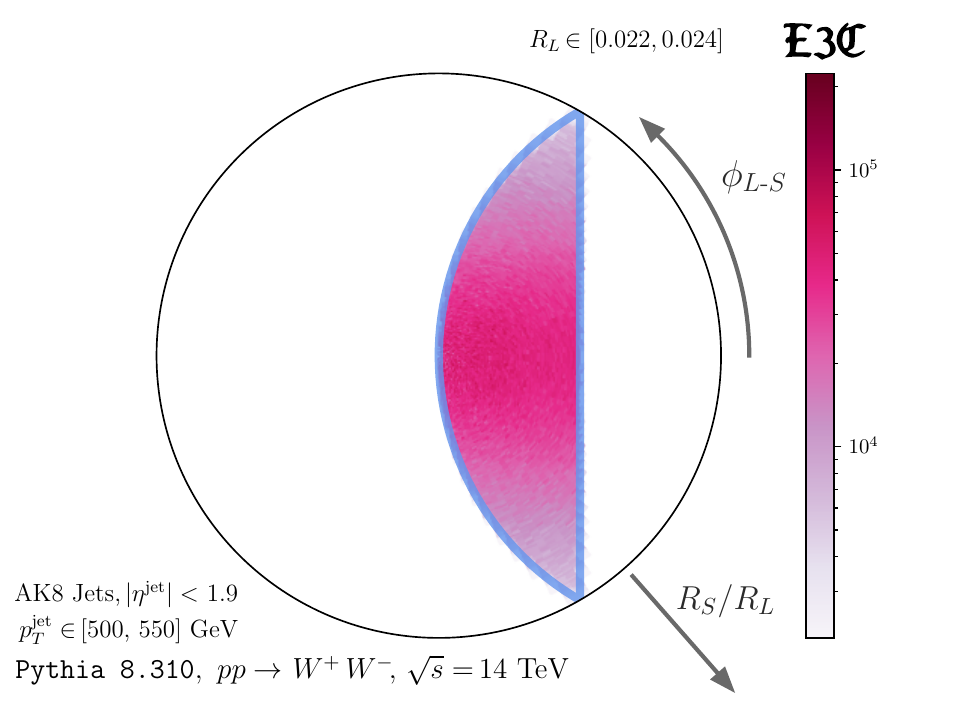}
    }
    \subfloat[]{
    	\includegraphics[width=0.34\textwidth]{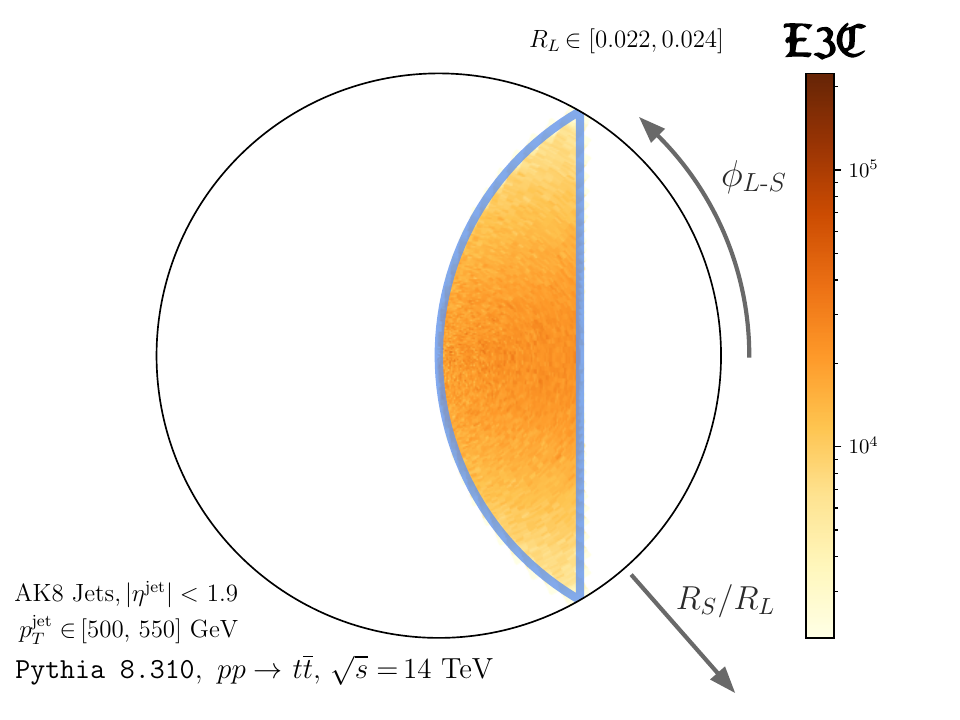}
    }
    \\
    \subfloat[]{
    	\includegraphics[width=0.34\textwidth]{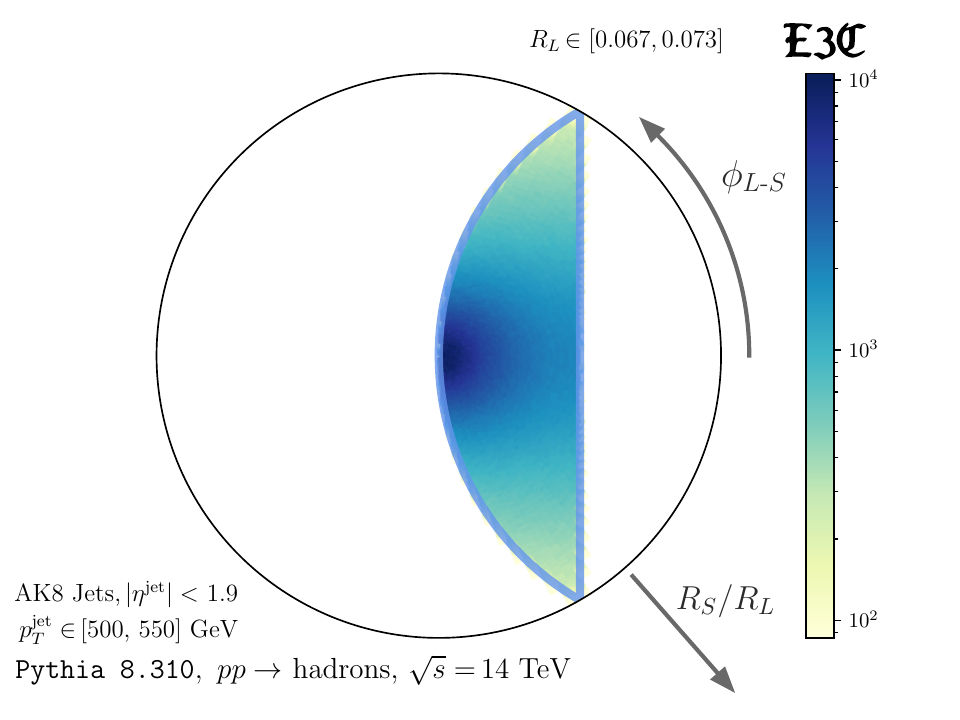}
    }
    \subfloat[]{
    	\includegraphics[width=0.34\textwidth]{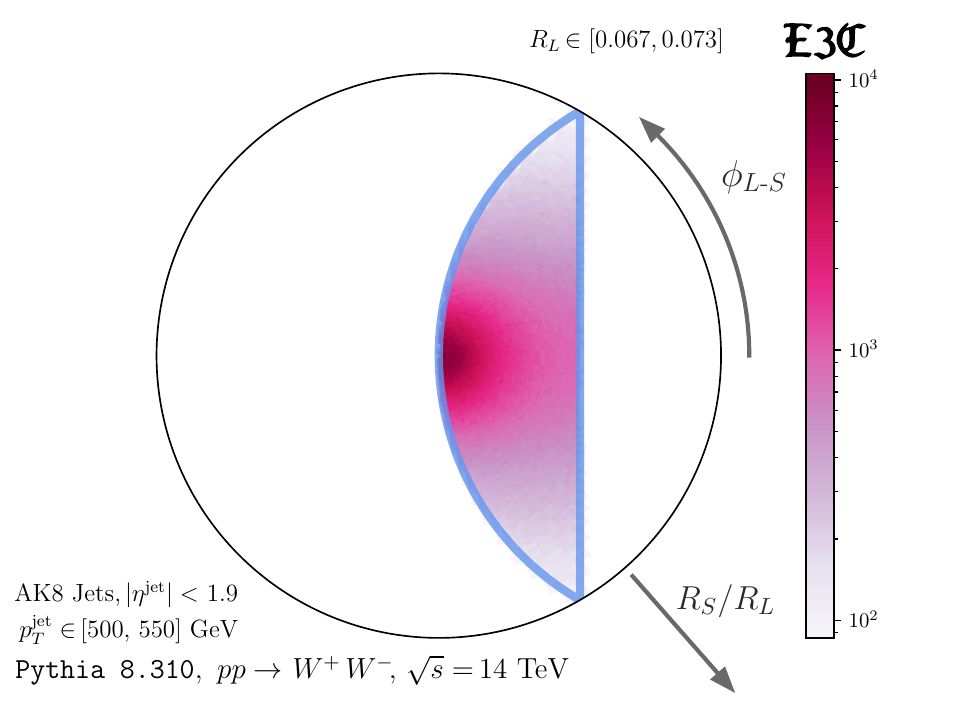}
    }
    \subfloat[]{
    	\includegraphics[width=0.34\textwidth]{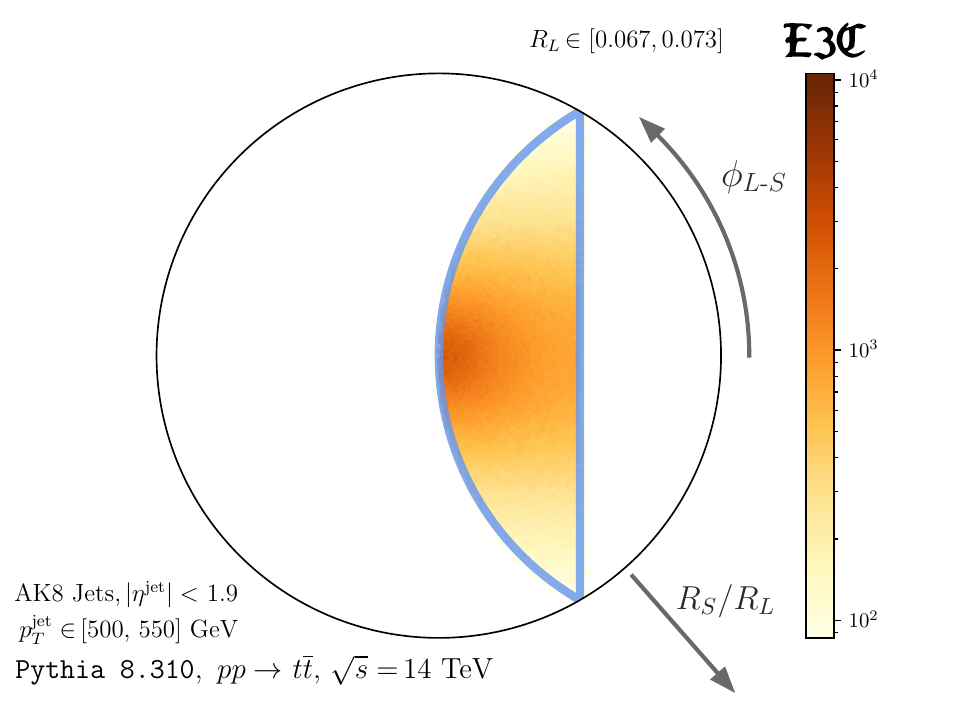}
    }
    \\
    \subfloat[]{
    	\includegraphics[width=0.34\textwidth]{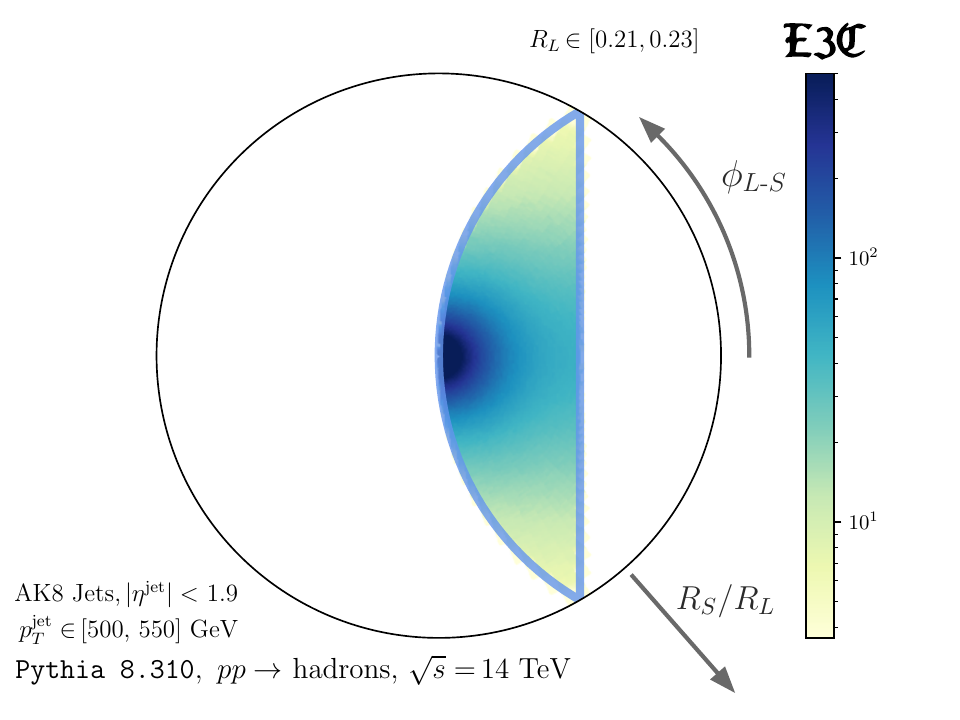}
    }
    \subfloat[]{
    	\includegraphics[width=0.34\textwidth]{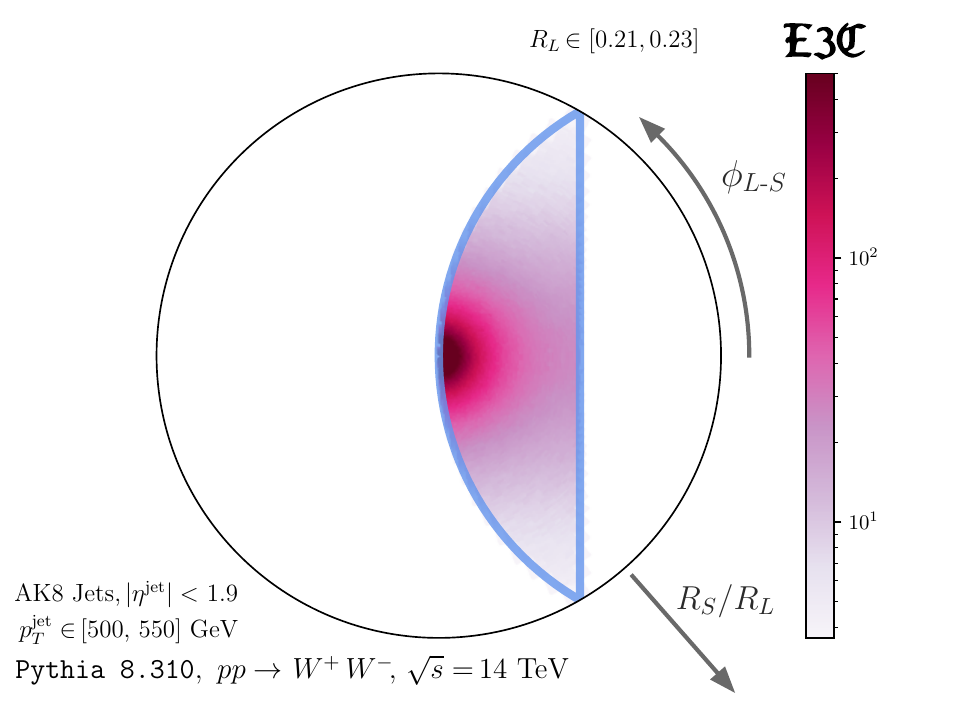}
        \label{fig:w_medium_trade3c}
    }
    \subfloat[]{
    	\includegraphics[width=0.34\textwidth]{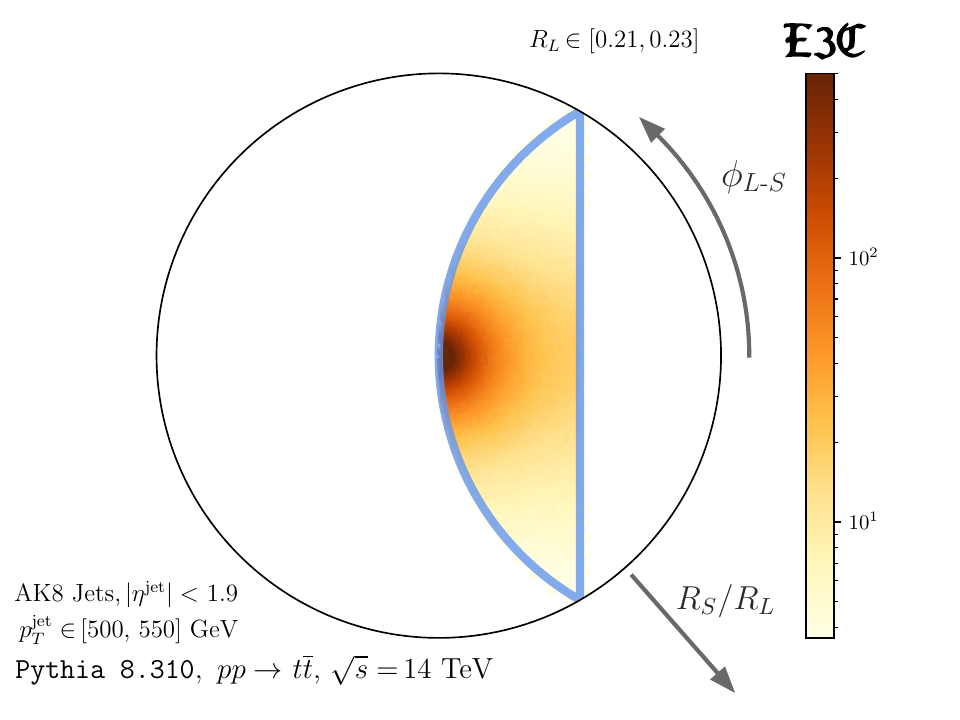}
    }
    \\
    \subfloat[]{
    	\includegraphics[width=0.34\textwidth]{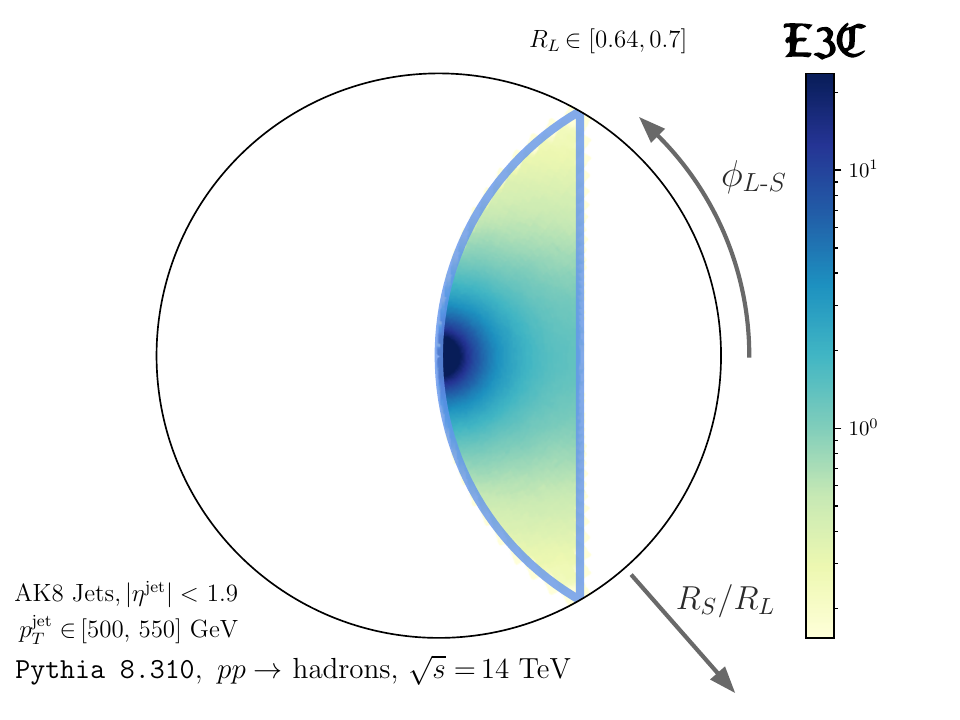}
    }
    \subfloat[]{
    	\includegraphics[width=0.34\textwidth]{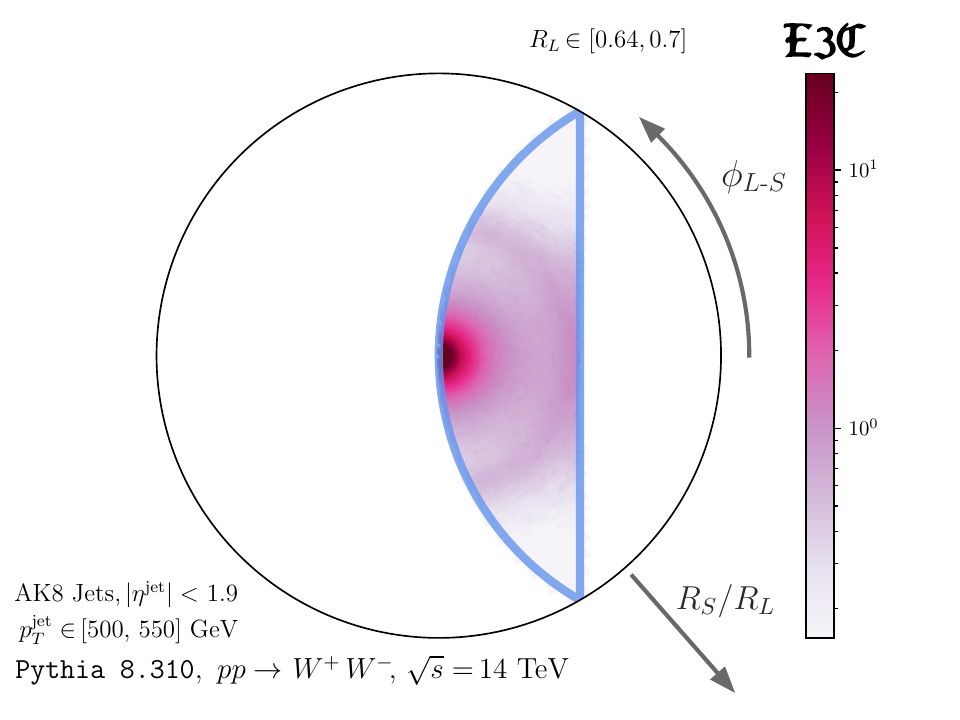}
        \label{fig:w_large_trade3c}
    }
    \subfloat[]{
    	\includegraphics[width=0.34\textwidth]{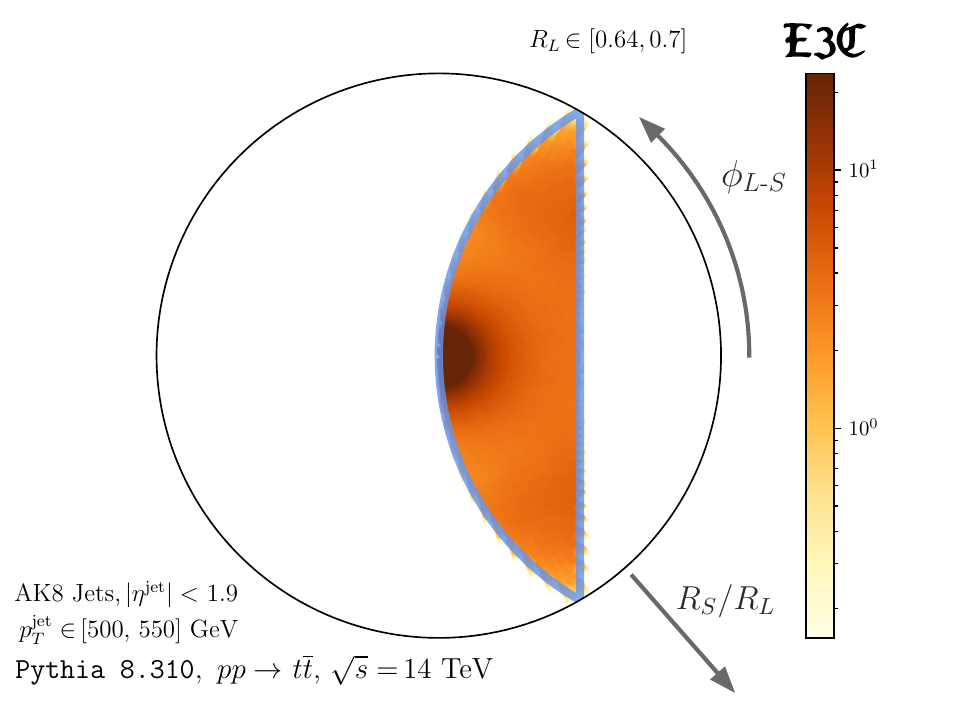}
    }
    \caption{
        Polar heat maps of the traditional E3C for (first column) QCD-, (second column) \(W\)-, and (third column) top-quark-initiated jets.
        The radial direction in each plot is the ratio \(R_S/R_L\), and the polar angle of the plot corresponds to the angle between the lines whose lengths determine \(R_S\) and \(R_L\).
        The \(R_L\) bins for each plot are chosen to be the same as the \(R_1\) bins in \fig{pythia_re3cs}.
        The polar heat maps of the traditional E3C contain similar information as the RE3Cs introduced in this work and shown in \fig{pythia_re3cs};
        however, the traditional E3C does not capture all the orientation information about radiation patterns in each jet sample that is present in our new RE3C.
    }
	\label{fig:pythia_old_e3cs}%
\end{figure}

The bullseye of representation of the RE3C we introduce are similar to the analogous bullseye representations of the traditional E3C for each jet sample, shown in \fig{pythia_old_e3cs}, for which the radial variable is the ratio \(R_S/R_L\) and the polar angle is the associated azimuthal angle.
However, the traditional E3C shows only a small slice of the information of the RE3C we introduce in this work, and the RE3C introduced in this work contains additional information about the relative orientations of particles within a jet;
for example, \figs{w_medium_re3c}{w_large_re3c} can be compared to \figs{w_medium_trade3c}{w_large_trade3c} to conclude that the RE3C we introduce conveys additional information about the orientation of the rings of radiation within \(W\)-boson-initiated jets.

\begin{figure}[t]
    \centering 
    \subfloat[]{
    	\includegraphics[width=0.34\textwidth]{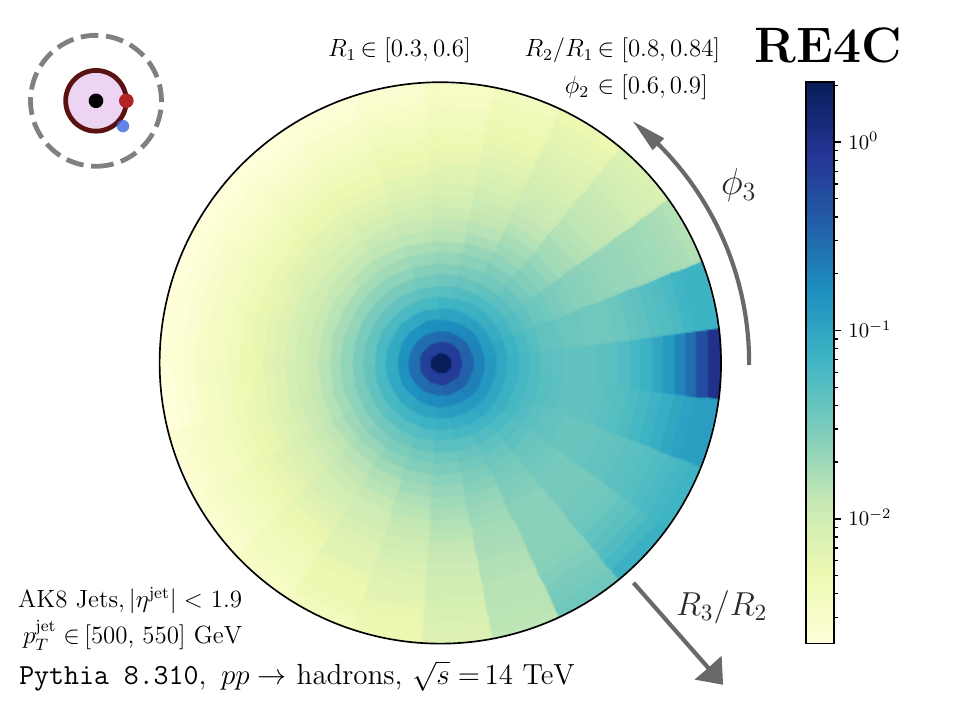}
    }
    \subfloat[]{
    	\includegraphics[width=0.34\textwidth]{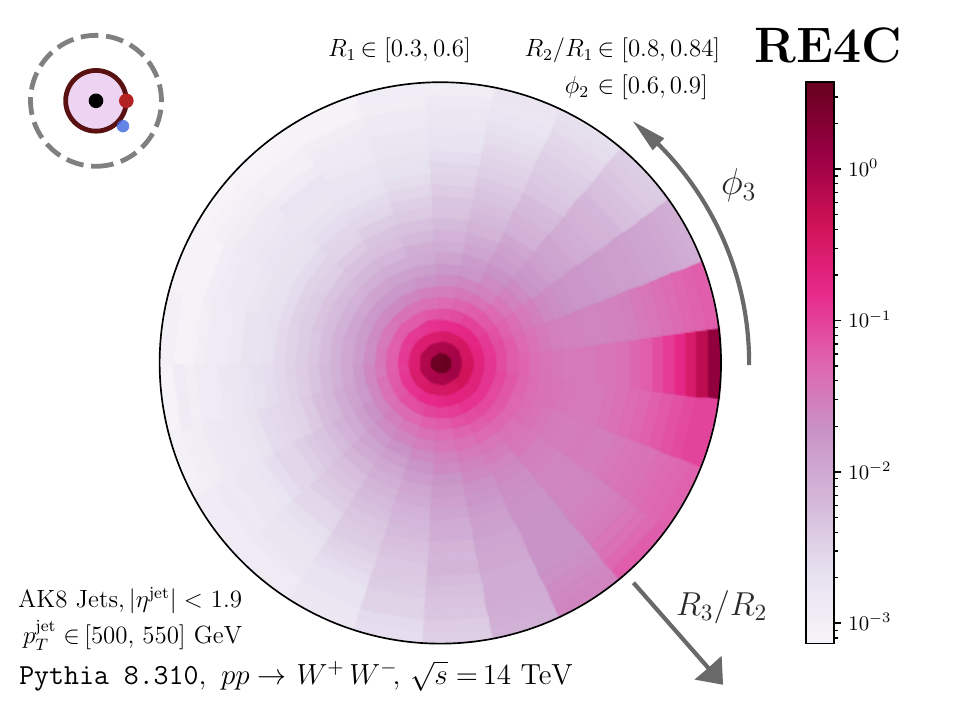}
    }
    \subfloat[]{
    	\includegraphics[width=0.34\textwidth]{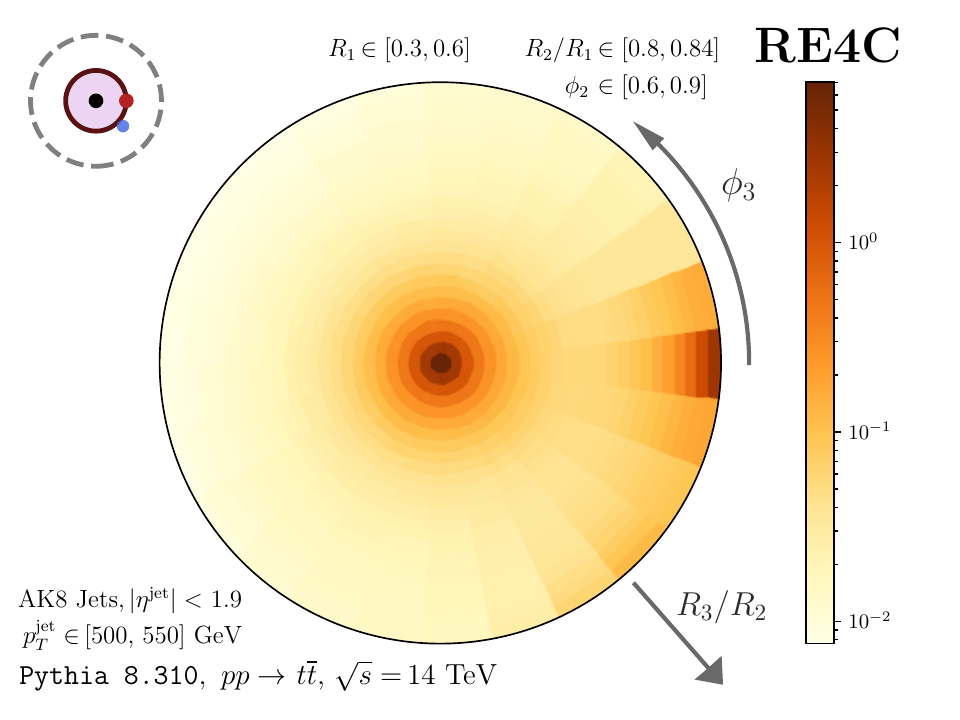}
    }
    \\
    \vspace{-8pt}
    \subfloat[]{
    	\includegraphics[width=0.34\textwidth]{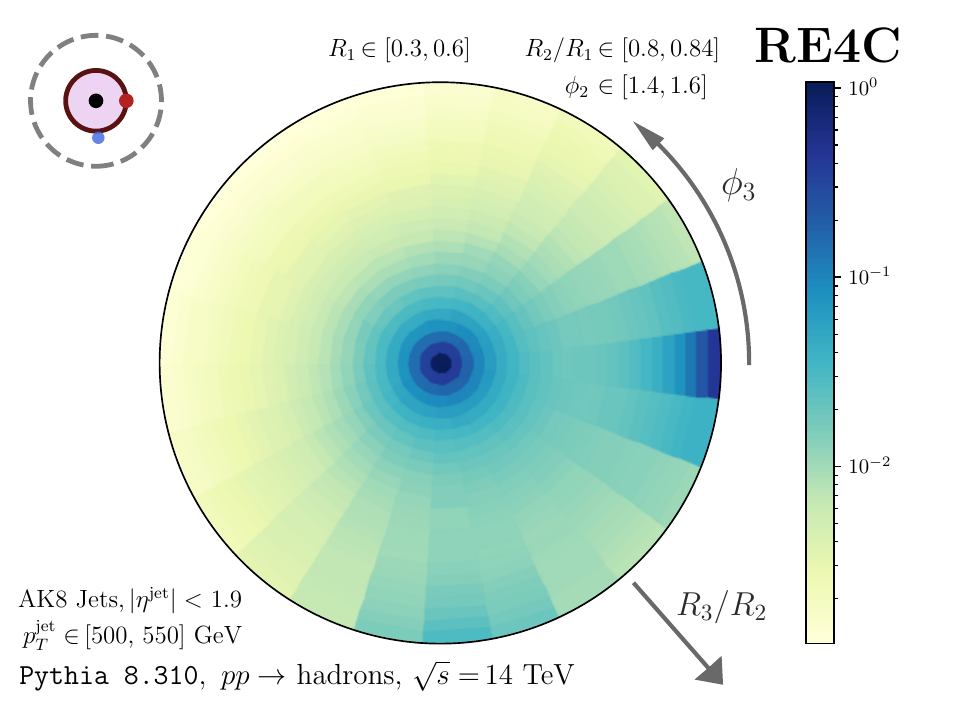}
    }
    \subfloat[]{
    	\includegraphics[width=0.34\textwidth]{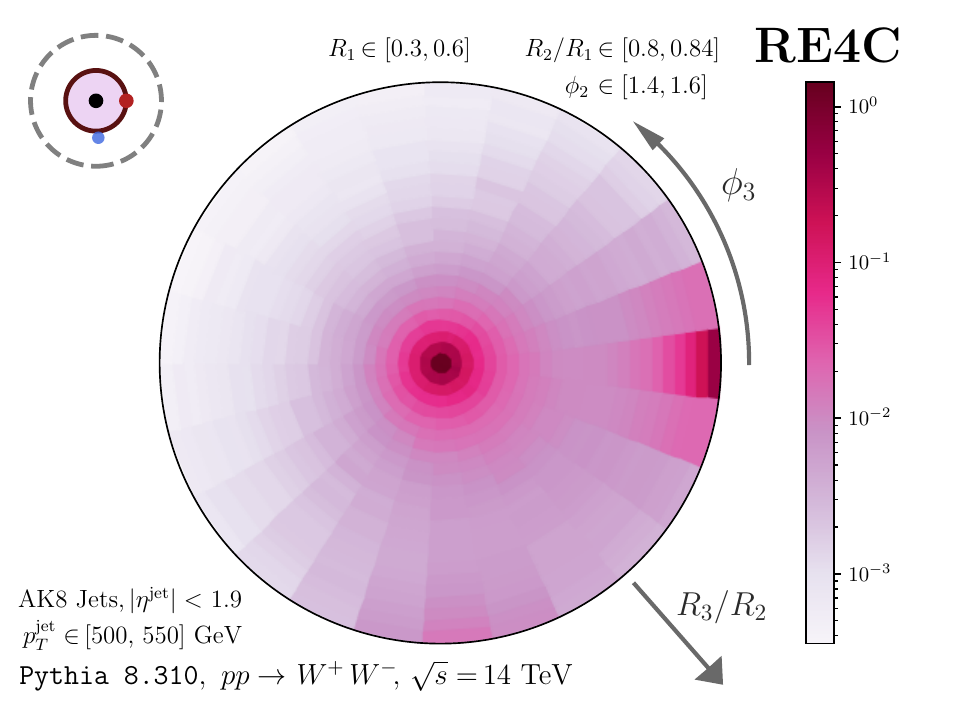}
    }
    \subfloat[]{
    	\includegraphics[width=0.34\textwidth]{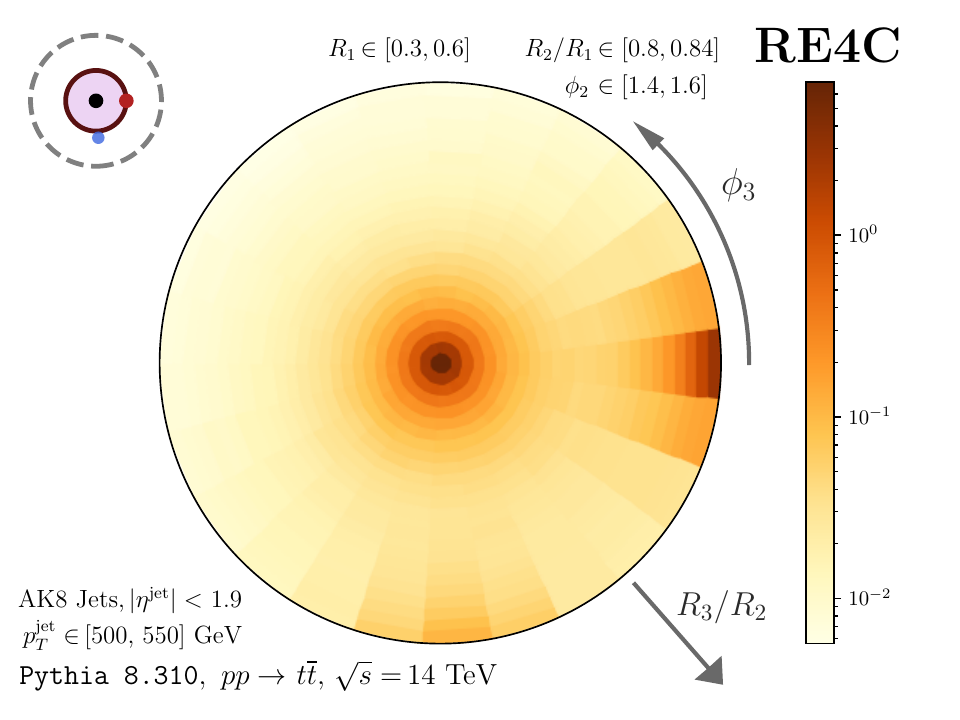}
    }
    \\
    \vspace{-8pt}
    \subfloat[]{
    	\includegraphics[width=0.34\textwidth]{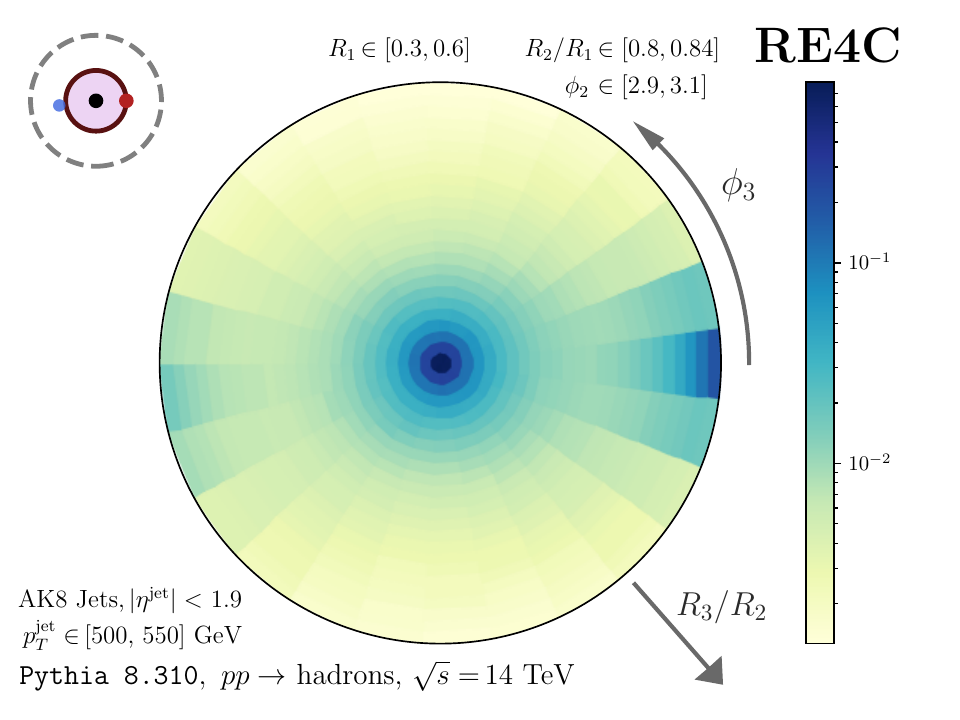}
    }
    \subfloat[]{
    	\includegraphics[width=0.34\textwidth]{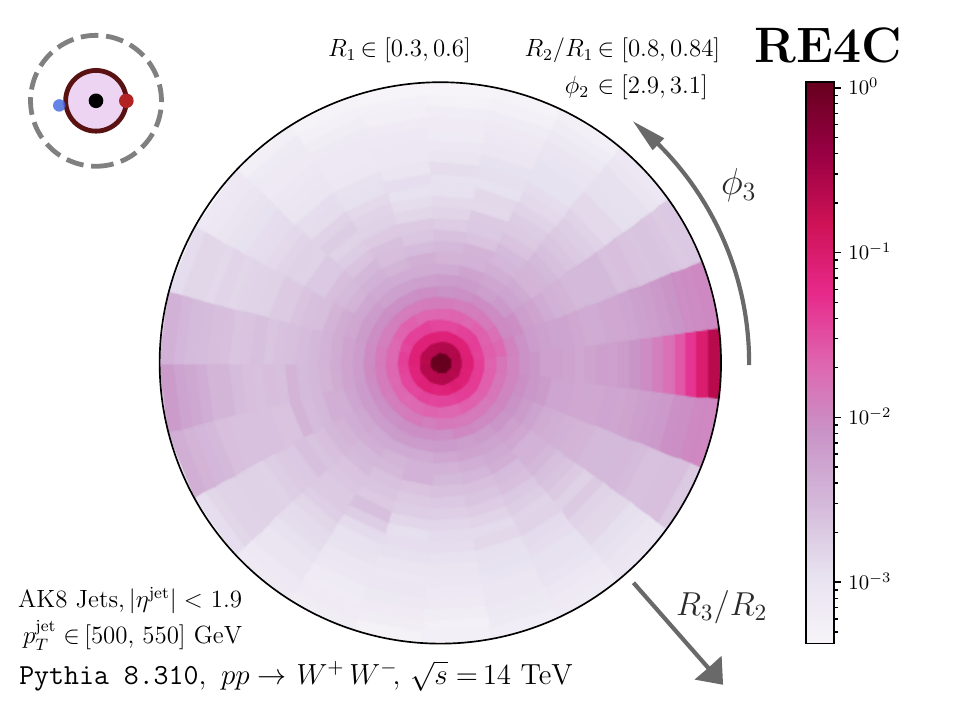}
    }
    \subfloat[]{
    	\includegraphics[width=0.34\textwidth]{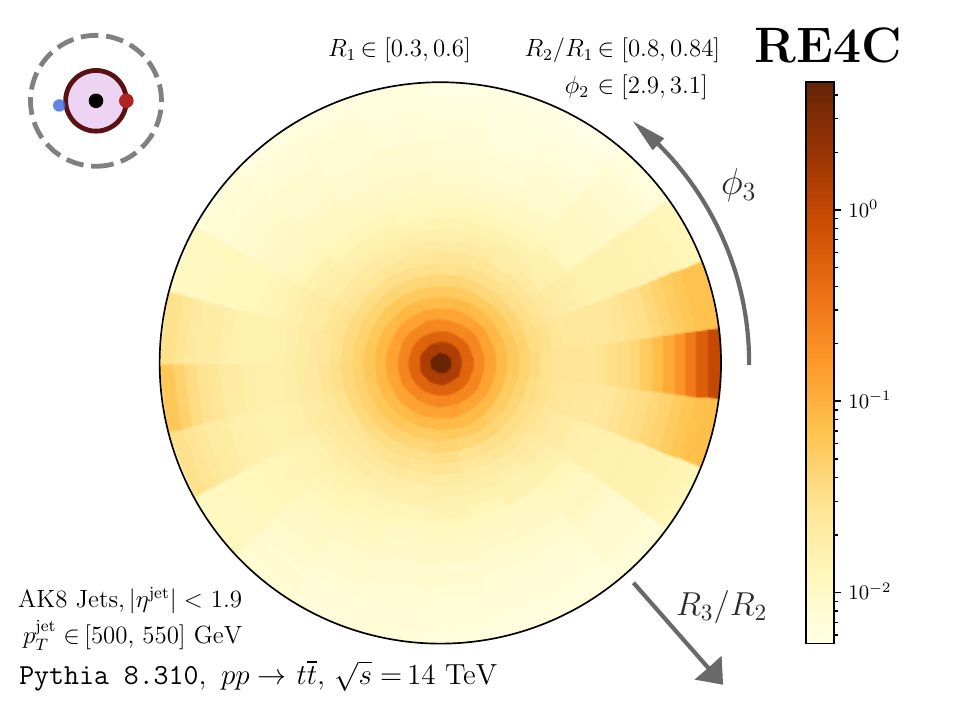}
    }
    \caption{
        Bullseye visualizations of the RE4C evaluated on QCD-, \(W\)-, and top-quark-initiated jets generated with \texttt{Pythia 8.310}, in the style of \fig{cms_re4cs}.
        As in \fig{cms_re4cs}, there are also enhanced correlations in energy when \(\phi_3\) is near \(-\phi_2\).
    }
	\label{fig:pythia_re4cs}%
\end{figure}

\fig{pythia_re4cs} provides similar polar heat maps for \(R_3/R_2\) and \(\phi_3\) for the RE4C, within several \(\phi_2\) bins, and for \(R_2\) near \(R_1\).
As in the case of the RE4Cs evaluated on CMS Open Data, we also see prominent correlations when \(\phi_3 \sim -\phi_2\) when \(R_2\) is near \(R_1\), corresponding to the case that particle \(i_3\) approaches particle \(i_1\).
These correlations for \(\phi_3 \sim -\phi_2\) quickly diminish as \(R_2\) is decreased and no longer near \(R_1\).

\begin{figure}[ht!]
    \centering
    \subfloat[]{
    	\includegraphics[width=0.34\textwidth]{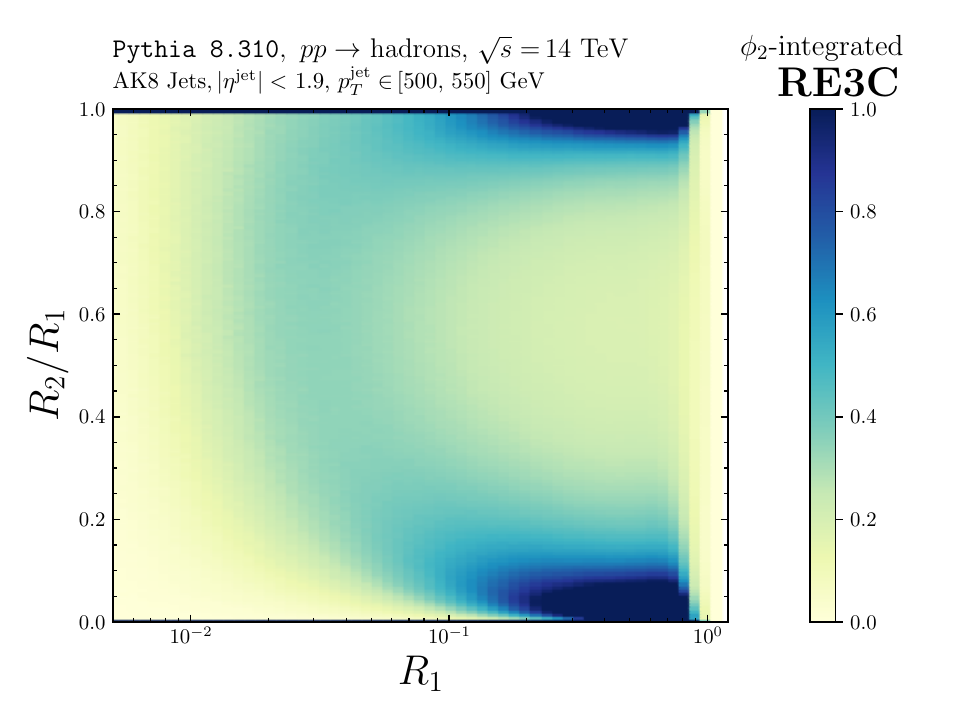}
    }
    \subfloat[]{
    	\includegraphics[width=0.34\textwidth]{figures/supplemental/density/w_newdef_density.pdf}
    }
    \subfloat[]{
    	\includegraphics[width=0.34\textwidth]{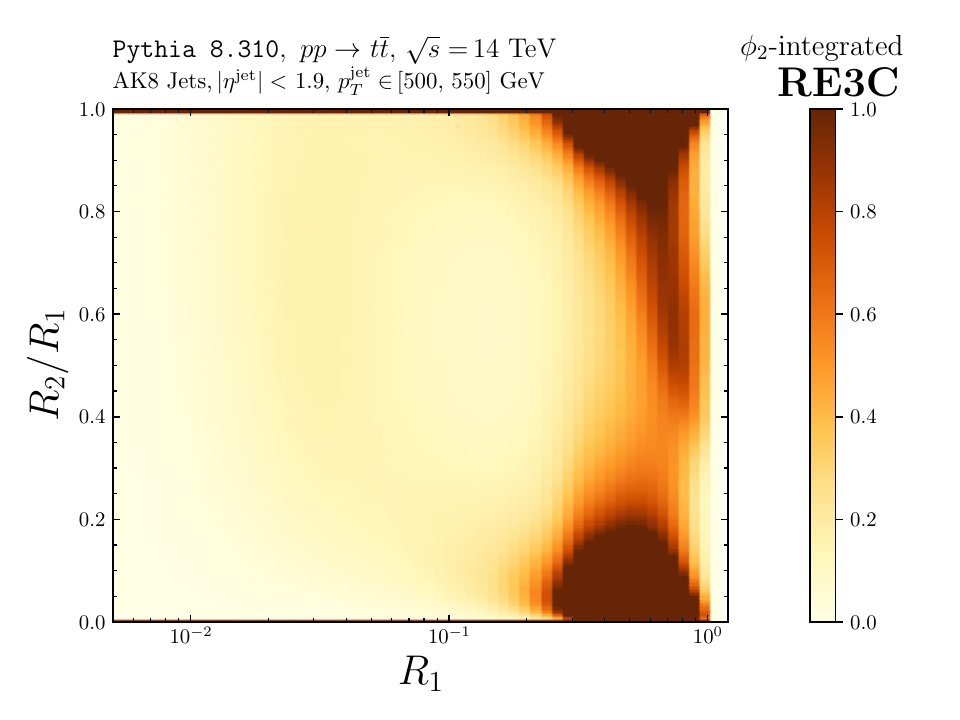}
    }
    \\
     \subfloat[]{
    	\includegraphics[width=0.34\textwidth]{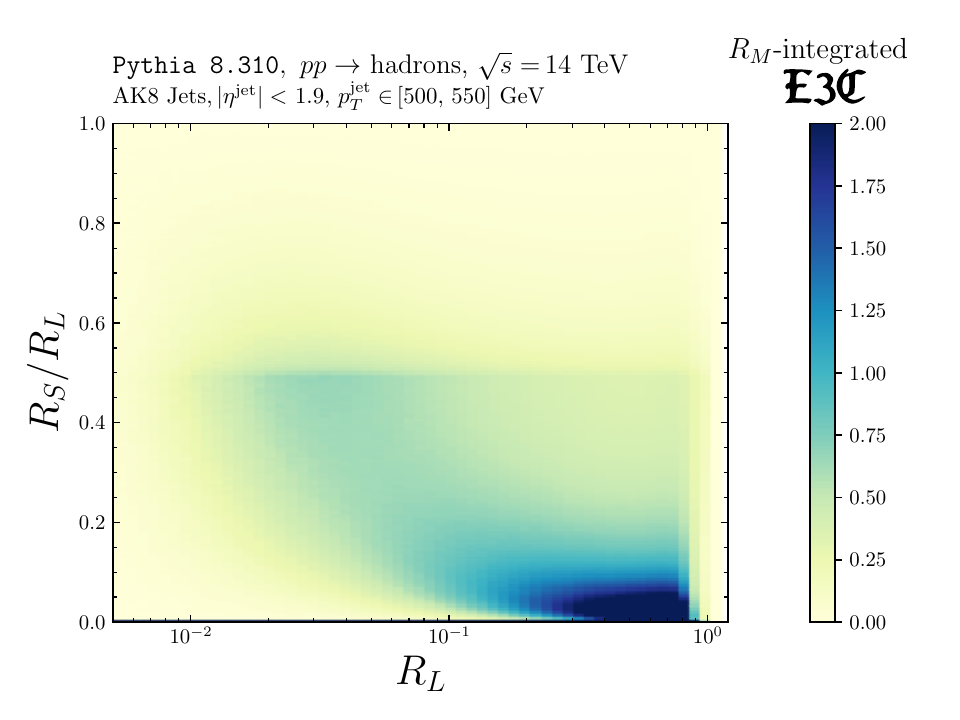}
    }
    \subfloat[]{
    	\includegraphics[width=0.34\textwidth]{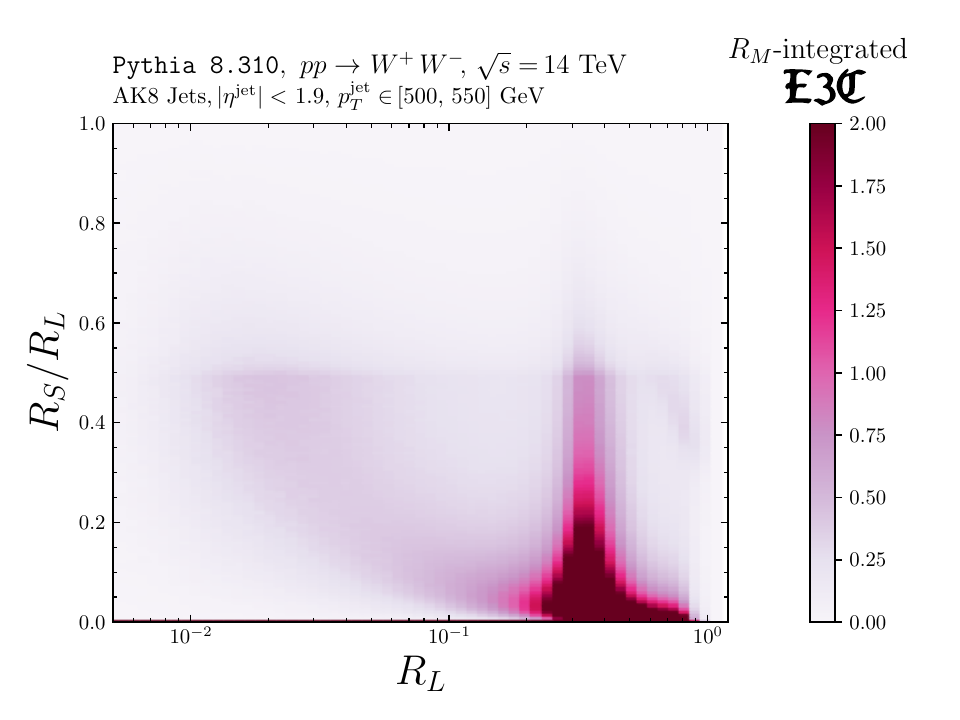}
    } 
    \subfloat[]{
    	\includegraphics[width=0.34\textwidth]{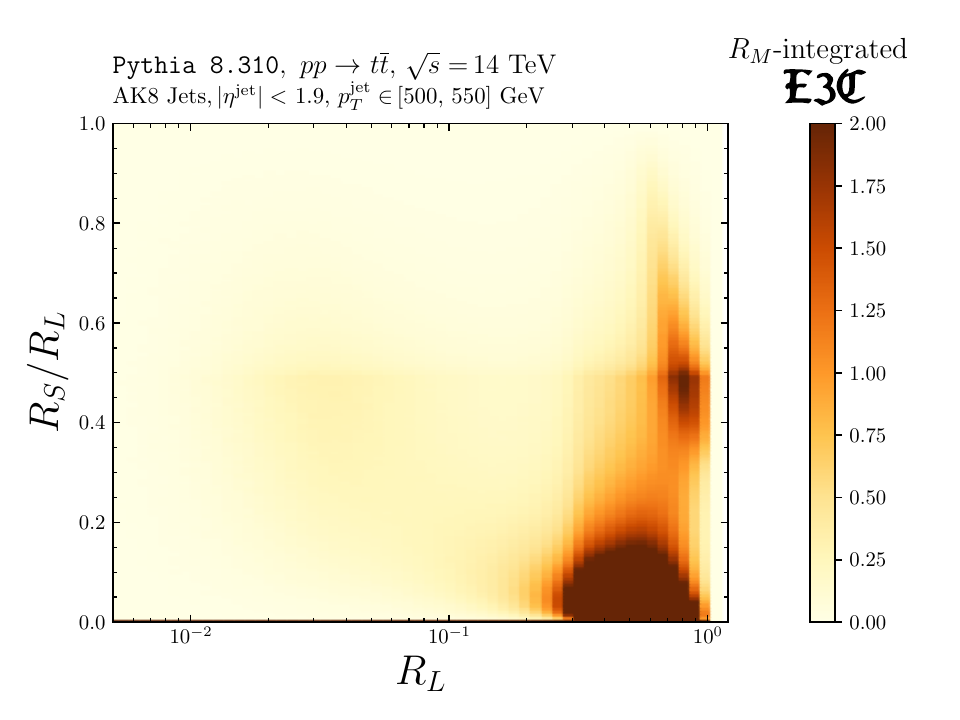}
    }
    \caption{
        Density plots for \(\phi_2\)-integrated RE3Cs (first row) and the analogous \(R_M\)-integrated traditional E3Cs (second row), in the style of Fig.~4 of the main text, evaluated on simulated events in \texttt{Pythia 8.310}.
        The first column shows \(\phi_2\)-integrated RE3Cs for \(p p \to \) hadrons, the second column for \(p p \to W^+ W^-\), and the third column for \(p p \to t \bar t\).
        Each row illuminates how three-point energy correlators encode the unique features of each type of jet.
        We note that the information of the traditional E3C is a subset of the information conveyed by the RE3C we introduce in this work.
    }
	\label{fig:pythia_densities}%
\end{figure}

Finally, we visualize the \(\phi_2\)-integrated RE3Cs for each jet sample in the first row of \fig{pythia_densities} and the \(R_M\)-integrated traditional E3Cs in the second row.
Much as in the case of the polar heat maps, the integrated RE3Cs densities provide strictly more information than the densities associated with traditional E3Cs. These provide a clear visual distinction between each sample of jets.

\section{Non-Perturbative Features}
\label{app:nonpert}

\begin{figure}[t]
    \centering 
    \subfloat[]{
    	\includegraphics[width=0.5\textwidth]{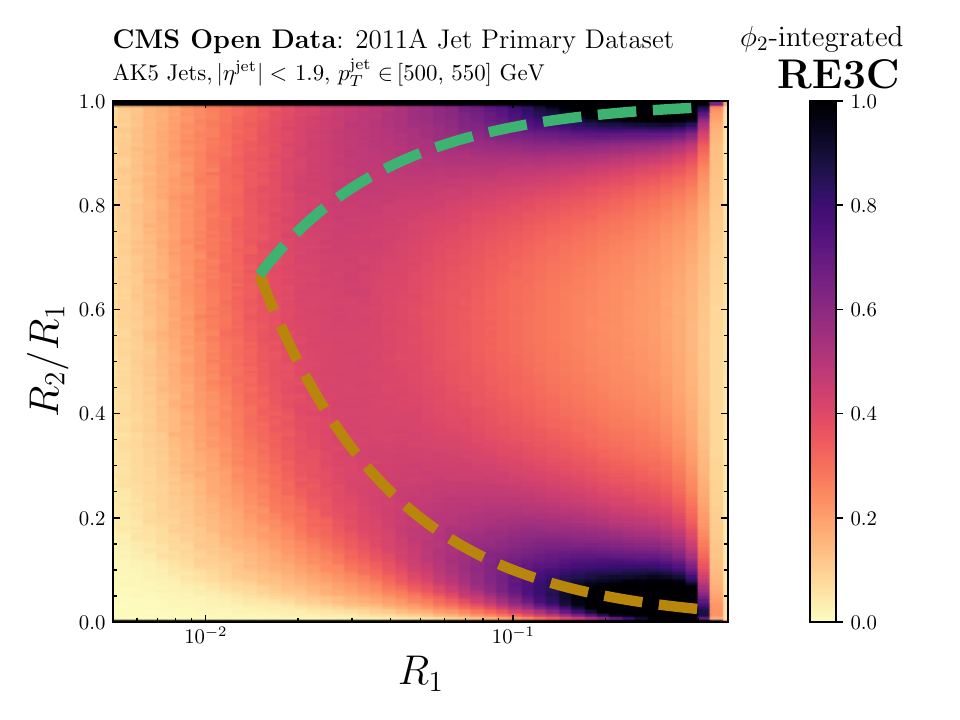}
    	\label{fig:nonpert_density}
    } 
    \subfloat[]{
        \begingroup
        \tikzset{every picture/.style={scale=1.3}}%
        	\begin{tikzpicture}
\definecolor{cornflowerblue}{rgb}{0.39, 0.58, 0.93}
\definecolor{azure(colorwheel)}{rgb}{0.0, 0.5, 1.0}

\definecolor{coralred}{rgb}{1.0, 0.25, 0.25}
\definecolor{cadmiumorange}{rgb}{0.93, 0.53, 0.18}
\definecolor{darkgoldenrod}{rgb}{0.72, 0.53, 0.04}
\definecolor{pastelorange}{rgb}{1.0, 0.7, 0.28}

\definecolor{rosevale}{rgb}{0.67, 0.31, 0.32}
\definecolor{palebrown}{rgb}{0.6, 0.46, 0.33}
\definecolor{goldenpoppy}{rgb}{0.99, 0.76, 0.0}
\definecolor{gold(metallic)}{rgb}{0.83, 0.69, 0.22}

\definecolor{heliotrope}{rgb}{0.87, 0.45, 1.0}
\definecolor{mediumorchid}{rgb}{0.73, 0.33, 0.83}

\definecolor{ao}{rgb}{0.0, 0.5, 0.0}
\definecolor{lightseagreen}{rgb}{0.13, 0.7, 0.67}
\definecolor{jade}{rgb}{0.0, 0.66, 0.42}

\colorlet{colsp}{ao}
\colorlet{coli1}{cornflowerblue}
\colorlet{coli2}{coralred}


\begin{scope}
    \clip(-1,-1.5) rectangle (5,2);
    \clip (0, 0) circle (5);

    \fill[heliotrope, opacity=0.4] (5, 0) circle (0.8);
    \fill[heliotrope, opacity=0.4] (0, 0) circle (0.8);

    \coordinate (O1) at (0, 0);
    \coordinate (E1) at (-35:0.8);
    \draw[|<->|,blue!50!red, very thick] (O1)--(E1)
    node[pos=0.8, right, yshift=-10pt, xshift=7pt, font=\Large]
    {\textcolor{blue!50!red}{$\frac{\Lambda_\text{QCD}}{p_T^\text{jet}}$}};
\end{scope}
\begin{scope}[xshift=5cm]
    \coordinate (O1) at (0, 0);
    \coordinate (E1) at (-140:0.8);
    \draw[|<->|,blue!50!red, very thick] (O1)--(E1)
    node[pos=0.8, right, yshift=-10pt, xshift=-38pt, font=\Large]
    {\textcolor{blue!50!red}{$\frac{\Lambda_\text{QCD}}{p_T^\text{jet}}$}};
\end{scope}

\draw[color=coli1, line width=0.5mm] 
    (0, 0) -- (0:5) coordinate (i1)
    node[pos=0.5, above, sloped, text=black,
         font=\Large]
        {\textcolor{coli1!50!black}{$\boldsymbol{R_1}$}};
    
\filldraw[color=coli1!50!black] 
    (i1) circle (3pt) 
    node[below right, font=\large]
    {\textcolor{coli1!50!black}{$i_1$}};

\draw[color=coli2, line width=0.5mm]
    (0, 0) -- (50:4.2) coordinate (i2)
    node[pos=0.6, left, text=black, xshift=-4pt, font=\Large]
    {\textcolor{coli2!50!black}{$\boldsymbol{R_2}$}};

\draw[-stealth, color=coli2, dashed, line width=0.4mm] 
    (0:4.2) 
    arc [start angle=0, end angle=48.5, radius=4.2]
    node[pos=0.5, right, text=black, font=\large]
    {\textcolor{coli2!50!black}{$\phi_2$}};

\filldraw[coli2!50!black] 
    (i2) circle (3pt)
    node[below, yshift=-5pt, font=\large]
    {\textcolor{coli2!50!black}{$i_2$}};

\filldraw[color=colsp!50!black]
    (0, 0) circle (3pt) 
    node[left, yshift=3pt, xshift=-5pt, font=\Large]
    {\textcolor{colsp!50!black}{$s$}};

\begin{scope}
    \clip (5, 0) circle (0.8);
    \draw[color=jade, 
    dash pattern=on 9pt off 3pt,
    line width=0.9mm, opacity=0.8] 
    (-30:4.65) 
    arc [start angle=-30, end angle=30, radius=4.65];
\end{scope}

\begin{scope}
    \draw[color=darkgoldenrod!95!black,
    dash pattern=on 10pt off 3pt,
    line width=0.9mm, opacity=0.8] (0, 0) circle (0.8);
\end{scope}

\end{tikzpicture}
        \endgroup
    	\label{fig:nonpert_cartoon}
    }
    \caption{
        Depictions of the non-perturbative features of the RE3C introduced in this work \textbf{(a)} highlighted in a two-dimensional density plot, and \textbf{(b)} in a simple cartoon demonstrating the relevant non-perturbative regimes.
        \textbf{(a)}
        A density plot for the RE3C evaluated on CMS Open Data and integrated over \(\phi_2\), taken from Fig.~4 of the main text, together with dashed lines indicating where non-perturbative effects are expected to become important using (geometric) phase space considerations.
        \textbf{(b)}
        The RE3C geometry, with non-perturbative regions of phase space highlighted in purple, and with the dashed lines near \(s\) and \(i_2\) indicating the values of \(R_2\) where we expect non-perturbative features to emerge in the integrated RE3C.
    }
	\label{fig:nonpert}
\end{figure}
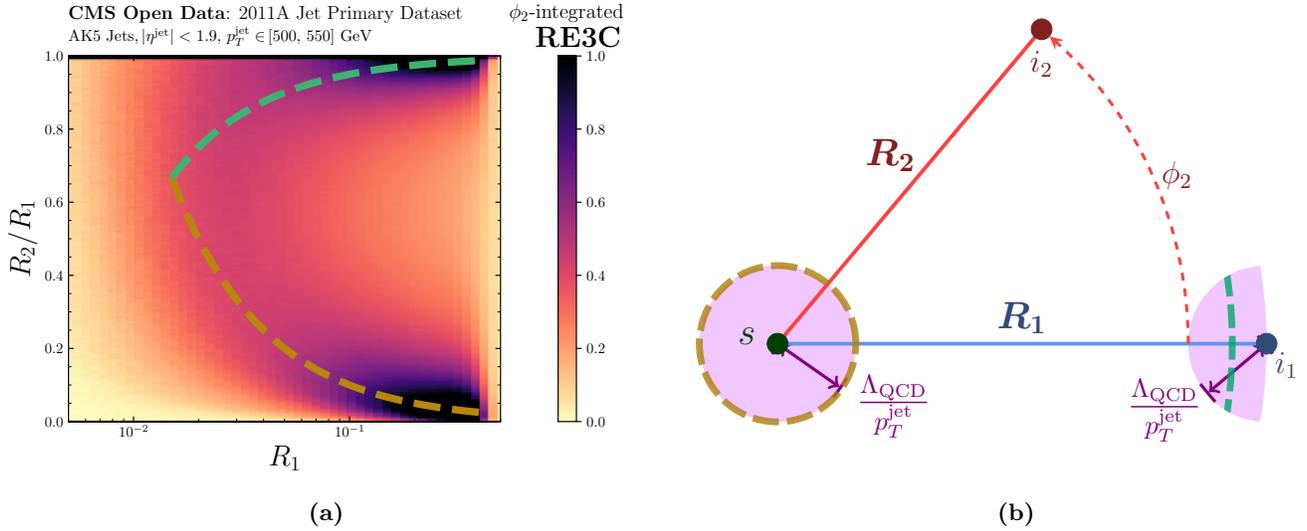

Finally, we investigate the non-perturbative features of the new RENCs we introduce in this work, focusing on the example provided by the density plots for the \(\phi_2\)-integrated RE3C which we visualize again in \fig{nonpert}.
For notational convenience in the discussion below, we let $p_T$ be shorthand for $p_T^\text{jet}$.
Unlike in the traditional E3C, for which there are only non-perturbative effects in the squeezed configuration with \(R_S \sim \Lambda_\text{QCD}/p_T\), our new RE3C exhibits non-perturbative features in two regimes:  (1) when \(R_2 \sim \Lambda_\text{QCD}/p_T\), when particle \(i_2\) (the emission that sets \(R_2\)) approaches the special particle \(s\); and (2) when \(R_1 - R_2 \sim \Lambda_\text{QCD}/p_T\), with the possibility that \(i_2\) approaches \(i_1\).

However, the shapes of the non-perturbative features of the integrated RE3C of \fig{nonpert_density} are not exactly symmetric under a reflection about the line \(R_2 = R_1/2\):
the ridge in \fig{nonpert_density} when \(R_2 \to R_1\) has a smaller slope than the ridge when \(R_2 \to 0\).
Concretely, this means that \(R_2\) must get closer to \(R_1\) in order for non-perturbative features to emerge than we might naively expect.
We refer to this as the ``\(R_2\)-asymmetry'' of the integrated RE3C.

To explain this \(R_2\)-asymmetry, we note that when $R_2 \sim \Lambda_{\rm QCD}/p_T$, shown by the dashed yellow line in \fig{nonpert}, the particle $i_2$ is non-perturbatively close to \(s\), independent of the orientation $\phi_2$ of particle $i_2$ around $s$.
On the other hand, when \(R_2 \sim R_1 - \Lambda_\text{QCD}/p_T\), particle \(i_2\) is not guaranteed to be near \(i_1\).
Instead, \(i_2\) is only non-perturbatively close to \(i_1\) in a narrow range of azimuthal angles around \(\phi_2 = 0\), as visualized in \fig{nonpert_cartoon}.
Therefore, it is only when \(R_2\) gets significantly closer to \(R_1\) that the non-perturbative features of the integrated RE3C begin to emerge.

We estimate the \(R_2\)-asymmetry of the integrated RE3C by noting that non-perturbative features emerge when the measure of the non-perturbative phase space for \(i_2\) near \(i_1\) is comparable to the non-perturbative phase space for \(i_2\) near \(s\).
This is visualized in \fig{nonpert_cartoon}:
non-perturbative features of the integrated RE3C emerge when the length of the green dashed line near \(i_1\) has a length of order \(\Lambda_\text{QCD}/p_T\).
The associated geometric constraint is a complicated solution to a cubic equation and we do not show it here.
However, we show our resulting estimate of the non-perturbative features as the green dashed line in \fig{nonpert_density}, which aligns closely with the non-perturbative ridge of the integrated RE3C as \(R_2\) approaches \(R_1\).

\pagebreak[4]

\end{document}